# HOUSING RISK AND RETURN: EVIDENCE FROM A HOUSING ASSET-PRICING MODEL

Karl Case, John Cotter, and Stuart Gabriel*
Wellesley College, University College Dublin, and UCLA

December 31 2010

## Abstract

This paper investigates the risk-return relationship in determination of housing asset pricing. In so doing, the paper evaluates behavioral hypotheses advanced by Case and Shiller (1988, 2002, 2009) in studies of boom and post-boom housing markets. The paper specifies and tests a multi-factor housing asset pricing model. In that model, we evaluate whether the market factor as well as other measures of risk, including idiosyncratic risk, momentum, and MSA size effects, have explanatory power for metropolitan-specific housing returns. Further, we test the robustness of the asset pricing results to inclusion of controls for socioeconomic variables commonly represented in the house price literature, including changes in employment, affordability, and foreclosure incidence. We find a sizable and statistically significant influence of the market factor on MSA house price returns. Moreover we show that market betas have varied substantially over time. Also, results are largely robust to the inclusion of other explanatory variables, including standard measures of risk and other housing market fundamentals. Additional tests of model validity using the Fama-MacBeth framework offer further strong support of a positive risk and return relationship in housing. Our findings are supportive of the application of a housing investment risk-return framework in explanation of variation in metro-area cross-section and time-series US house price returns. Further, results strongly corroborate Case-Shiller survey research indicating the importance of speculative forces in the determination of U.S. housing returns.

*Case is Hepburn Professor of Economics, Department of Economics, Wellesley College, Wellesley, Massachusetts, e-mail: kcase@wellesley.edu; Cotter is Associate Professor of Finance, Director of Centre for Financial Markets, UCD School of Business, University College Dublin, Blackrock, Co. Dublin, Ireland, email john.cotter@ucd.ie and Research Fellow, Ziman Center for Real Estate, UCLA Anderson School of Management. Gabriel is Arden Realty Chair and Professor of Finance, Anderson School of Management, University of California, Los Angeles, 110 Westwood Plaza C412, Los Angeles, California 90095-1481, email: stuart.gabriel@anderson.ucla.edu. This research was funded by the UCLA Ziman Center for Real Estate. Cotter acknowledges the support of Science Foundation Ireland under Grant Number 08/SRC/FM1389. The authors thank Xudong An, Michael Brennan, Jan Brueckner, Steve Cauley, Jerry Coakley, Joao Cocco, Will Goetzmann, Lu Han, Matt Kahn, Stuart Myers, Maureen O' Hara, Robert Shiller, Tsur Sommerville, Richard Roll, Matthew Spiegel, Bill Wheaton and participants at the 2010 American Real Estate and Urban Economics Society Annual Conference, the 56th Annual Meetings of the Regional Science Association International, 2010 UCI-UCLA-USC Research Symposium, and seminar participants at the Hebrew University of Jerusalem, University of New South Wales, University of Melbourne and University College Dublin for comments The authors are grateful to Ryan Vaughn for excellent research assistance.

## 1. Introduction

The boom and bust of house prices defined the opening decade of the 21st century. Excessive movements in house prices figured importantly in the 2007 meltdown in mortgage and capital markets and the downturn in the global economy. Further, neither analysts on Wall St., regulators in Washington, D.C., nor academic economists well anticipated the depth of the house price cycle, its geographic dispersion, or related asset class contagion. While speculative motives were evidenced in recent homebuying behavior, prevailing house price models instead largely focused on long-run fundamentals (e.g., Case and Shiller (1988, 1990), Case and Quigley (1991), Gabriel, Mattey and Wascher (1999), Himmelberg, Mayer, and Sinai (2005)). However, ongoing behavioral research by Case and Shiller (1988, 2002, 2009) suggests that fundamentals are inadequate to explain house price fluctuations and that speculation plays an important role.

This paper introduces a risk-return framework in examination of the role of speculation in housing investment returns. We do so via development and test of a multi-factor asset pricing model for housing (see, for example, Fama and French, 1992).[1] The analysis first examines the role of market returns as a common factor and evaluates the suitability of alternative market proxies. The research then seeks to determine whether additional measures of house price risk, including momentum, volatility, and geographic arbitrage between high- and low-priced areas have explanatory power for housing returns.[2] We also assess the robustness of the risk-return relationship in the presence of non-risk characteristics, including employment, affordability, and foreclosure effects. We examine the relation both in the cross-section and time-series of metropolitan house price returns. In this manner, the research seeks new insights as regards the extreme movements in house prices evidenced in many U.S. metropolitan markets over recent decades.

[1] As is well known, the most commonly examined risk-return relationship is the CAPM, whereby an asset's or portfolio's returns are predicted by the market portfolio return. This model is typically applied to the pricing of equities where the market portfolio return is proxied by an equity index or some other diversified portfolio of equities. A criticism is of this approach is that the market portfolio cannot be proxied by a restrictive set of assets such as that contained in an equity index, and as a consequence, the model cannot be adequately tested (Roll, 1977). Work has been done to develop more comprehensive market portfolios including proxies for returns to human capital (Campbell, 1996); however, even those more comprehensive measures of market return still exclude many assets, most notably the returns to housing investment, the largest element of household wealth. That not withstanding, it remains standard practice in equity pricing to use equity index values to proxy for market prices, and thereby, to estimate market returns. Here an index is formed through a weighting (e.g., value-weighting) of individual equities. This analysis adopts an analogous approach in assessment of housing asset pricing. We use house price indexes derived from individual housing transactions; also, we apply anaggregated national series to proxy the market return. It is also standard practice in equity pricing to use indexes, for example the S&P500, to represent all assets in the economy notwithstanding the Roll critique. We do recognise that our use of a national housing series has a similar limitation in representing all assets in the economy. Accordingly and as an alternative, we also examine the use of the S&P500 as a proxy for market returns. Note that the legitimacy of explaining an asset's return with a single market variable also has been questioned (Fama and French, 1996). In defence, however, the market factor has been found to be the most important factor that predicts equity returns. These limitations in the pricing and testing framework have led to the development and application of multi-factor models, most notably the Fama-French 3 factors (Fama and French, 1992).

[2] Housing is analogous to equities in that it can pay two forms of compensation to investors. For equities, compensation is composed of price returns and dividends, whereas for housing compensation is comprised of house price returns and rents. Similar to the standard approach taken in the equity pricing literature, (e.g. Fama and French, 1993) we focus on modelling only the price return compensation of housing investment. Furthermore, in keeping with the strategy followed in the equity pricing literature, we recognize that while the assumptions of the application of factor models

Results of estimation of a single market factor housing model provide evidence of a strong positive relationship between housing risk and returns. This relationship is robust to the inclusion of controls for well-known economic fundamentals. Using the Fama-MacBeth (1973) framework to test the pricing model, we find strong support for the basic premise of the single factor model, that there is a positive risk and return relationship in housing asset pricing. We also find evidence of non-linearity in the beta risk and return relationship. Moreover, the Fama-MacBeth analysis indicates that housing investors seek compensation for total risk, including both market and idiosyncratic locational risk.

Our approach stands in marked contrast to much of the literature in housing economics. Often, those models pool cross-location and time-series data in reduced form specifications of supply- and demand-side fundamentals, including controls for labor market, nominal affordability, and other cyclical terms (see, for example, Case and Shiller (1988, 1990), Case and Quigley (1991), Gabriel, Mattey and Wascher (1999), Himmelberg, Mayer, and Sinai (2005)). While house price determination has been a popular topic of economics research, (see, for example, Case and Shiller (1989, 2003)), existing models often have failed to capture the substantial time and place variability in housing returns.

Note also that ongoing behavioral research by Case and Shiller (1988, 2003, 2009)) suggests a market for residential real estate that is very different from the one traditionally discussed and modeled in the housing economics literature.[3] Early surveys of recent homebuyers in San Francisco, Los Angeles, Boston, and Milwaukee, Case and Shiller (1988) concluded that "without question, home buyers [in all four sampled areas] looked at their decision to buy as an investment decision."[4] More recent survey findings point to the growing importance of investment motivations for home purchase. For example, results of the 2002 survey, published in Case, Quigley, and Shiller (2003), indicated that investment returns were an important consideration for the vast majority of home buyers. Further, the pattern of survey

---

may not fully hold for equity investment, and similarly for investment in housing, this does not invalidate testing the appropriateness of these models for housing.

[3] Case and Shiller (1988) conclude that "In a fully rational market, prices would be driven by fundamentals such as income, demographic changes, national economic conditions and so forth. The survey results presented here and actual price behavior together sketches a very different picture. While the evidence is circumstantial, and we can only offer conjectures, we see a market driven largely by [investment] expectations." Further, speculation may be reinforced and augmented by money illusion (Brunnermeier and Julliard, 2008) where investors see house price increases in nominal terms, and fail to see them in real house price changes.

[4]Case and Shiller (1988) note that "home buyers in the boom cities had much higher expectations for future price increases, and were more influenced by investment motives. In both California cities, over 95 percent said that they thought of their purchase as an investment at least in part. In Boston, the figure was 93.0 percent. A surprisingly large number in San Francisco, 37.2 percent, said that they bought the property "strictly" for investment purposes."

findings reveals both geographic and temporal variations in investment demand for homeownership. In discussion of recently released 2009 Case-Shiller survey results (see New York Times, October 11, 2009) Bob Shiller suggests that "the sudden turn in the housing market probably reflects a new homebuyer emphasis on market timing." Shiller concludes that "it appears the extreme ups and downs of the housing market have turned many Americans into housing speculators."[5]

To assess the dynamics underpinning house price returns, we specify and test a multi-factor housing asset pricing model.[6] We assume that the investment decision is restricted to housing. This strategy is consistent with the extensive literature in equity asset pricing, where investment in that asset class is assumed to be segmented, rather than integrated.[7] Despite the fundamental importance of factor models to empirical asset pricing (see, for example, Fama and MacBeth (1973), Merton (1973), Fama and French (1992), Fama and French (1993), Roll (1977)), few papers have undertaken comprehensive tests of the investment asset pricing framework in applications to housing.[8] In this paper, we assess the importance of the market return (as proxied by aggregate US housing returns) to expected returns in metropolitan housing markets. However, as described below, a first consideration is to assess the appropriateness of alternative proxies for the market factor, including both the national house price series and the S&P500 equity return series.

Moreover we augment and develop a multi-factor model by examining the impact of other risk factors including idiosyncratic risk, momentum, and house price size effects as are commonly cited in the equity pricing literature. An extensive debate has focused on the validity of market returns alone explaining the variation in expected returns. That debate has resulted in the development of multi-factor models, for example, Arbitrage Pricing Theory (APT). These models support the inclusion of additional factors and we follow this approach in determining whether additional factors help to explain the variation in expected house price returns.

[5] Case, Quigley, and Shiller (2003) suggest that even after a long boom, home-buyers typically had expectations that prices over the next 10 years would show double-digit annual price growth, apparently only with a modest level of risk. Results from 2008 and 2009 Case and Shiller surveys provide strong evidence that homebuyers remain housing bulls in the long-run. Further, they suggest that "it seems reasonable to conjecture that an expectations formation process such as this could well be a major contributor to the substantial swings seen in housing prices in some US regions."

[6] This paper should be seen as a distinct approach to the consumption based asset pricing models for housing (see Lustig and Van Nieuwerburgh, 2005 and Piazzesi, Schneider, and Tuzel, 2007; and Han, 2009).

[7] We recognize the value of developing a more comprehensive benchmark portfolio that may include investment in housing, equities and human capital. These are usually not pursued in the literature (an exception being a portfolio compromised of equities and human capital (Campbell, 1996)) due to issues such as weighting structure and data availability.

[8]While homeownership user cost computations account for expected housing investment returns, standard reduced form house price models focus largely on fundamentals associated with housing consumption demand.

Idiosyncratic risk would not be included in the traditional single factor model as market risk is taken to be the sole predictor of expected returns. In that context, investors are assumed to hold a fully diversified market portfolio. However, investment in housing typically is not associated with large-scale diversification, as investors typically hold a small number of location-specific properties (for example, a single property) in private ownership. This suggests that the housing pricing model should not only include a reward in expected returns for systematic (market) risk, but also provide compensation for diversifiable risk.[9] Thus, housing investors seek compensation for total risk, encompassing both systematic (market) risk and unsystematic (idiosyncratic) risk (see Merton's (1987) model for a theoretical framework). In the empirical asset pricing literature, however, evidence on the role of idiosyncratic risk for equity pricing is mixed. Ang, Hodrick Xing and Zhang (2006) find the relationship between idiosyncratic risk and expected returns is negative. In contrast, Goyal and Santa-Clara (2003) find a positive relationship, whereas Bali, Cakici, Yan, Zhang (2001) find an insignificant relationship. For real estate, the issue is largely overlooked, although Plazzi, Torous and Valkonov (2008) find a positive relationship between commercial real estate expected returns and idiosyncratic risk.[10] We use the most commonly applied measure of idiosyncratic risk by taking the standard deviations of the squared residuals from the single market factor model. Regardless, idiosyncratic risk is an important component of total risk for equities (Campbell, Latteau, Makiel, Xu, 2000) and given a lack of diversification may also be prominent for housing investment.[11]

Also in the equity pricing literature, research has confirmed the existence of a size effect whereby small firms earn higher risk-adjusted returns than large firms (using firm market capitalization as a measure of firm size). Banz (1981) was among the first to document the size effect--suggesting that returns on small firms were high relative to their betas. The prevalence of this effect led Fama and French (1992) to incorporate size as a risk factor in the multi-factor framework. Known as Small Minus Big (SMB), this control tests for a zero cost investment strategy based on size whereby investors short large firms to finance their ownership of small firms. Fama and French (1992) find a positive relationship between the SMB factor and expected returns and show that it predicts future asset returns. In housing research, Cannon, Miller and Pandher (2007) find a positive cross-sectional relationship between the SMB factor and housing returns. We construct a similar SMB term for metropolitan housing by subtracting the $75^{th}$ quartile return based on median MSA house prices from the $25^{th}$ quartile return for each time interval.

Carhart (1997) has provided evidence in support of the inclusion of a momentum term in the pricing of equities. The momentum term seeks to identify past winners and losers in asset returns and specifies a trading strategy by assuming that these outcomes will continue in the future. In that trading strategy, the investor buys past winners and sells past losers with the expectation that the overall return is positive.[12] In a key study, Jegadeesh and Titman (1993) sort past returns into decile portfolios and assume the investor buys the best return ranking portfolio and sells the worst return ranking portfolio for each period. The authors find that their momentum factor has significant positive explanatory power for equity returns, and remains even in the presence of the control for market risk. In addition, an extensive literature has used variations on this definition with similar results. Momentum has been generally overlooked in the housing literature although momentum trading has been found to have a positive influence on future real estate investment trust (REIT) returns (Chui, Titman, and Wei, 2003; Derwall, Huij, Brounen, and Marquering, 2009). For our asset pricing model, winning and losing MSAs are identified in every time period by sorting all previous period's MSA returns and the highest (lowest) returns are associated with winners (losers). In specification of this housing spatial arbitrage term, we take an average of the lagged highest decile returns less an average of the lagged lowest decile returns.

[9] Of course idiosyncratic risk may also have an influence on house prices for different reasons. For instance if there is mispricing of housing it will attract economic agents such as arbitragers who try and exploit this and earn non-market risk related returns.

[10] The mixed evidence may result from the modelling of idiosyncratic risk where a number of alternative measures are driven by different econometric assumptions (eg. see Lehmann, 1990).

[11] As in the case of equities, idiosyncratic risk associated with housing investment may have changed over time. For example, as shown by Campbell, Latteau, Makiel, and Xu (2000), idiosyncratic risk trended upwards up during the 1990s, but this trend has reversed in more recent times (Bekaert Hodrick Zhang, 2008).

Finally, we assess the robustness of the augmented asset pricing model results to the inclusion of controls for local economic fundamentals commonly represented in the house price literature. Glaeser and Gyourko (2007), for example, note that most variation in house price returns is local. Accordingly, model controls seek to link house price fluctuations to local fundamentals including nominal ability-to-pay, employment changes, and foreclosure incidence (see, for example, Case and Shiller (1988, 1990), Goodman and Gabriel (1996), Case and Quigley (1991), Gabriel, Mattey and Wascher (1999), Himmelberg, Mayer, and Sinai (2005) and Glaeser and Gyourko (2007).

Our focus on the cross-sectional and intertemporal dynamics of US house prices is facilitated via the application of quarterly house price indices from the Office of Federal Housing Enterprise Oversight (OFHEO) for the 1985-2007 timeframe and across over 150 MSAs.'[13] We also confirm that model results are robust to use of the Case-Shiller repeat sales house price indexes; which are available for a limited number of US cities. The OFHEO metropolitan area series offer a substantially greater number of observations than the Case-Shiller data, but unlike the Case-Shiller series, are limited to sales and refinancings using conventional, GSE-conforming loans. The national house price series is identified as the market return for housing investment.[14] The study first uses a pooled cross-section and time-series approach to fit the asset pricing model. We generate betas for each MSA's returns with respect to movements in the OFHEO national house price index. Each beta represents the market risk-adjusted sensitivity of the per-period change in MSA-specific house prices to movements in the aggregate housing market. High betas represent high risk housing markets whereas low betas represent low risk housing markets. For example, as expected, we find high housing betas in metropolitan areas of the east and west costs, notably including California and Florida, whereas areas of the upper mid-west and Great Plains are characterized by low betas. In general, we find that investment in high (low) risk markets is compensated by high (low) returns.

---

[12] Anecdotal reports during the boom period suggested the increased prevalence of housing investment trades across geographies. For example, during that period, housing in Las Vegas was viewed as substantially more affordable than that in Los Angeles. Moreover, housing in Los Vegas often also exhibited higher returns than that of Los Angeles. Media reports documented investors selling their homes in higher-priced and lower-return Los Angeles, and buying in lower-priced and higher-return Los Vegas (for example, see Annette Haddad, *Los Angeles Times*, 2006; who details housing investors who used profits from sales of homes in LA to invest in Arizona and Las Vegas).

[13] Note that the Housing and Economic Recovery Act of 2008 created a new mortgage market regulator known as the Federal Housing Finance Agency (FHFA), which merged the activities of OFHEO, the the Federal Housing Finance Board (FHFB), and the GSE mission office at the Department of Housing and Urban Development (HUD). The OFHEO data can now be found at the FHFA website.

[14] In contrast aggregate stock market returns have a negligible influence on the variation of house price returns with low explanatory power, and is supportive of previous evidence (Case, 2000).

We also undertake cross-sectional analysis at quarterly intervals for our large sample of MSAs to examine the temporal evolution of our asset pricing variables. Assessment of the time-series of our model coefficients indicates that the relative importance of explanatory factors has varied across time and over the housing cycle. Specifically we find that the positive influence of the market factor on MSA-specific asset returns has been marked by substantial cyclical variability in some metropolitan areas; in other areas, betas have evidenced little increase or decrease. Further, as expected, the model explanatory power does vary substantially across MSAs, suggesting the housing investment framework is more relevant to an explanation of house price returns in some MSAs, whereas in other places, housing largely remains a consumption good. To illustrate, we find that market betas increase substantially through the sample period for Milwaukee, where those estimates are estimated at close to zero through much of the 1990s, but then rise to about 1 toward the end of the sample period. In contrast, the opposite occurs in Boston, where market betas are estimated at greater than 2 early in the time-series but trend down to less than 1 during the mid-2000s, only to jump again precipitously during the subsequent housing boom years. However there are a large set of MSAs where the market betas remain relatively high or low throughout the sample. Also the asset pricing model explanatory power varies across MSAs; for example, model fit is higher for coastal areas such as Washington, D.C. ($R^2 = 0.75$). As expected, in other areas, including many small Midwestern metropolitan areas, the asset pricing investment model does not capture variation in the house price data ($R^2 = 0$ in Cedar Rapids, IA).

We also run separate time-series models for each MSA. We find strong evidence of a risk-return relationship that varies across MSAs. In particular our market betas vary substantially and are strongly related to the relative explanatory power of the models in the cross-section. The average correlation across MSAs between the $R^2$ and betas for our housing asset pricing model with only 1 factor, the OFHEO National series, is 0.739. In terms of specific MSAs we find that Raleigh-Cary, NC has a very low explanatory power ($R^2 = 0.018$) coupled with a low beta (0.135) whereas in contrast Tampa-St.Petersburg-Clearwater, FL has a relatively high $R^2$ (0.668) and market beta (1.8).

To avoid a potential errors-in-variable problem from using single assets, we also examine the pricing relationship using portfolios of MSA returns. Using portfolios we test the validity of our housing asset pricing model using the Fama-MacBeth (1973) framework. Note, however, that using portfolios is not without its challenges. Roll (1977) finds that portfolio averages may conceal relevant information on assets, so as to make it difficult to determine the impact of variables on asset returns.[15] This issue is particularly relevant to studies of metropolitan housing markets (relative to equity markets), in that limited cross-sectional housing data may give rise to portfolios containing few assets. That notwithstanding, we find a strong risk and return relationship for the housing portfolios. Our findings corroborate survey findings by Case and Shiller and are supportive of the application of a housing investment risk-return model in explanation of variation in metro-area cross-section and time-series of US house price returns. Further, our results suggest the markedly elevated importance of a housing investment asset pricing framework to certain MSAs over the course of the recent house price cycle.

[15] Also the portfolio sort criteria has an impact on the findings for portfolio returns with Brennan, Chordia and Subrahmanyam (1998) showing that the impact on returns change significantly from using 6 versus 7 portfolios.

The plan of the paper is as follows. The following section describes our house price data and characterizes temporal and cross-sectional variability in house price returns. Section 2.2 defines model explanatory variables and reports on summary characteristics in the data. Section 2.3 reports on the estimation results of alternative specifications of the housing asset pricing model, inclusive of assessment of cross-sectional and temporal variation in the housing market betas. Section 2.4 focuses on model validation using Fama-MacBeth analysis, followed by robustness checks in Section 2.5. Section 3 provides concluding remarks.

## 2. Analysis

### 2.1 Housing Market Returns

In our asset pricing model the dependent inputs include MSA-specific house price returns as proxied by the OFHEO metropolitan indices. Regression analysis is undertaken on 151 MSAs for which we obtained quarterly price index data from 1985:Q1 – 2007:Q4. The house price time-series are produced by the U.S. Office of Federal Housing Enterprise Oversight (OFHEO). The OFHEO series are weighted repeat-sale price indices associated with single-family homes. Home sales and refinancing activity included in the OFHEO sample derive from conventional home purchase mortgage loans conforming to the underwriting requirements of the housing Government Sponsored Enterprises—the Federal National Mortgage Association (Fannie Mae) and the Federal Home Loan Mortgage Corporation (Freddie Mac). The OFHEO data comprise the most extensive cross-sectional and time-series set of quality-adjusted house price indices available in the United States. However, due to exclusion of sales and refinancing associated with U. S. Government (FHA and VA) and non-conforming home mortgages, the OFHEO series likely understates the actual level of geographic and time-series variability in U.S. house prices.[16]

While some of the MSA-specific OFHEO series are available from 1975, our timeframe (1985-2007) is chosen so as to maximize representation of U.S. metropolitan areas.[17] Our 151 time-series include all major U.S. markets. OFHEO actually provides data for a larger number of MSAs (384 in total for 2009) which is used to create the National house price index. However, many of those MSAs are associated with a lack of trading activity and so the full set of MSAs are not included as rankable according to the definition provided by

[16] For a full discussion of the OFHEO house price index, see "A Comparison of House Price Measures", Mimeo, Freddie Mac, February 28, 2008.

[17] The Case-Shiller house price indices provide the primary alternative to the OFHEO series. While the Case-Shiller price indices are not confined only to conforming mortgage transactions, they include a substantially smaller (N=16) set of cities beginning from 1990. We repeat our analysis using the Case-Shiller cities and also present these results.

OFHEO. Moreover our sample is restricted to include only those MSAs with data available between 1985 and 2007 resulting in 151 individual MSAs. However we are confident that we have captured a very large proportion of US housing market as measured by OFHEO with the average of individual MSA series very strongly correlated with the National series (corr = 0.953). We calculate house price returns for each MSA in our sample as the log quarterly difference in its repeat home sales price index.[18]

Figure 1 provides an initial review of the house price series incorporating time series plots and summary details at quarterly frequency. Here, for illustrative purposes, we distinguish movements in house prices for the 4 metropolitan areas identified in ongoing Case-Shiller survey research, relative to that of the U.S. market overall. As suggested above, the OFHEO national series is computed over a large number of sampled areas for the 1985-Q1 through 2007-Q4 period. In each case, the time-series of index levels are normalized to 100 in Q1 1995.

Just in these cities alone, figure 1 provides evidence of considerable temporal and cross-sectional variation in the house price series. As shown, the rate of increase in aggregate market returns accelerated markedly during the post-recession years of the early 2000s. Among the 4 identified locations, extreme house price run-ups are identified for coastal metropolitan areas, with the highest rates of mean price change and risk (standard deviation of index changes) shown for California coastal markets. In Los Angeles, for example, house prices moved up from an index level of 100 in 1995 to a peak level of almost 350 in 2007! One quarter’s returns almost reached 10%. Similar price movements, although somewhat less extreme, were evidenced in San Francisco and Boston. In marked contrast, house price trend and risk were substantially muted in Milwaukee, at levels close to the US market average.

## 2.2 Inputs to the Regressions

### Explanatory Variables

Table 1 provides definitions and summary information on model variables. These include standard measures of risk as well as additional local economics controls. While empirical modelling is undertaken at a quarterly frequency, the summary statistics of model variables are displayed at an annual frequency. As shown, the time-series average return for all MSA housing markets ($R_{HPI}$) is positive and substantial at almost 1% per annum with an average deviation of about 0.74%. Moreover we see strong temporal variation with returns ranging from -0.295% to 2.530%. This is similar to the national OFHEO series ($R_{OFHEO}$). The alternative market return series, the S&P 500 ($R_{SP}$), is characterized by substantially elevated risk relative to that of housing markets and relatively poor returns.

We also seek evidence of elevated returns among lower-priced metropolitan areas. In the context of housing, the small minus big term (SMB) is defined as the quarterly return associated with the 25$^{th}$ percentile house price MSA less that associated with the 75$^{th}$ percentile house price MSA. As suggested above, SMB has been found to be an important determinant of equity returns, as small (market capitalization) firms earn higher returns than large firms (see, for example, Banz, 1981 and Fama and French, 1992). For US housing markets, the average SMB return is a positive 0.175, moreover, SMB does exhibit substantial variation and is more than 2 standard errors from zero ($t = 0.175/(0.406/\sqrt{23})$).

[18] In principle, it would be desirable to model house prices at higher frequencies. Unfortunately, monthly quality-adjusted house price indices are available from OFHEO only for Census Divisions (N=18) and only for a much shorter time-series.

As in the equity asset pricing literature, idiosyncratic risk ($s^2$) is defined as the standard deviation of squared asset pricing model residuals (see Ang, Hodrick, Xing and Zhang, 2006). Accordingly, $s^2$ provides a proxy for diversifiable risk. In marked contrast to equities, a typical housing investor trades in a very small number of location-specific properties, suggesting that diversification in housing investment is substantially more difficult to achieve. Again, relative to equities, idiosyncratic risk should be relatively more important to housing investment (as has been found by Plazzi, Torous and Valkanov (2008) in the case of commercial real estate). As shown in Table 1, we find substantial idiosyncratic risk on average (4.590%) that is 4.86 standard errors from zero ($t = 4.59/(4.53/\sqrt{23})$), with considerable temporal variation in this variable. Idiosyncratic risk is also heavily right skewed as suggested by the median mean relation.

Consistent with the finance literature (e.g. Jegadeesh and Titman, 1993), our momentum term reflects average house price return differentials between the lagged 10 highest and lowest return sample MSAs for each quarter. This formulation tests the hypothesis that investors identify the best performing MSAs in the country and fund investments in those areas via sales of property in the worst performing areas. The average return from the momentum strategy is large (6.350%) and is statistically greater than zero ($t = 10.26$). Accordingly, the momentum term seeks to identify speculative spatial strategies among housing investors.

The final three variables, quarterly proxies for change in employment ($\Delta$Emp), change in foreclosures ($\Delta$Forc)), and log of lagged affordability, (log($Afford_{t-1}$), are economic factors commonly cited in the housing literature. In that regard, nominal affordability is particularly important to mortgage qualification and related demand for housing. Further, as suggested by the above citations, housing returns are taken to vary with fluctuations in local employment and foreclosure activity. As indicated in Table 1, all terms are presented at yearly frequency. The employment variable represents the one quarter log change in MSA employment using data supplied by the Federal Reserve Bank of St. Louis. On average, employment fell by about 0.7 among MSAs in the sample. Affordability is defined as the log of the one quarter lagged ratio of MSA mean household income to mean house price. In our sample, housing affordability averaged 0.241% and is statistically significant in 47 of the 151 MSAs. Foreclosure information is provided by the Mortgage Bankers Association and is defined as the 1-quarter change in foreclosures per MSA. Foreclosures are substantial and average over 1% per MSA. These levels are significant across housing markets.

Table 2 provides a matrix of simple correlations among the time-varying variables. As evident, there exists little correlation between the housing market ($R_{HPI}$) and equity return ($R_{SP}$) series. In marked contrast, and as would be expected, the correlation between the MSA cross-sectional average housing market return ($R_{HPI}$) and that of the OFHEO index ($R_{OFHEO}$) exceeds 0.95. As evaluated below, the Table is suggestive of the importance of the national housing return series ($R_{OFHEO}$) in determination of returns at the MSA level ($R_{HPI}$). The Table further reveals a relatively strong correlation between the housing market return series ($R_{HPI}$) and the Small minus Big (SMB) term. Otherwise, simple correlations with the remaining explanatory variables are of limited magnitude with the exception of the affordability and foreclosures terms. Generally we also note a lack of correlation between the explanatory variables, suggesting we can isolate the impact of these variables on the variation of house prices.

## Estimating Housing Market βs

### 2.3 β Estimates

Table 3 presents results of our factor asset pricing models. The table provides summary evidence on regressions estimated for each of the 151 MSAs included in the analysis. For each explanatory variable, Table 3 presents the average estimated coefficient value. The number of MSAs with significant estimated coefficients is indicated in parentheses below the coefficient values. The models' slope coefficients and associated $R^2$ provide evidence on the ability of risk factors and controls to explain variation in house price returns. Models (1) – (6) present variants of the basic model; those specifications are indicated in a memo item to the table. In addition, the tables provide additional summary information based on estimation results for the 151 MSAs on model coefficients and model explanatory power. Model (1) consists of the single market factor housing model; here we regress returns in each MSA ($R_{HPI}$) on national housing market returns ($R_{OFHEO}$). In model (2), we estimate an alternative single market factor housing model, whereby a proxy for equity market returns ($R_{SP}$) is used to represent the market variable.[19]

We separately generate betas for each MSA with respect to movements in the market return. Each beta represents the sensitivity of the quarterly change in the MSA-specific OFHEO index to movements in the specified market factor. High betas represent high risk MSA housing markets, whereas low betas represent the opposite. In the basic pricing framework of model (1), a MSA's quality-adjusted house price returns are generated by market risk only. In equity markets, the market factor is typically proxied by a broad market portfolio such as the S&P500. We examine two alternative proxies of the market factor, the log difference of the OFHEO national house price index, and the log difference of the S&P500 index, both at quarterly frequency. The OFHEO national series is an equally weighted index of the individual MSA house price indices, whereas the stock market index is a value-weighted series.[20]

To begin, we identify the relevant market factor as the National OFHEO series. As shown in model (1) for the 151 MSAs sampled, the average estimated market beta is close to 0.8; further, the housing market return proxy is statistically significant in 103 MSAs. Note also that the mean $R^2$ in the OFHEO series single factor model is almost 20 percent. Those results stand in marked contrast to findings associated with the equity market return series. Results from model (2) indicate the lack of power or significance of the equity market return series in explaining MSA-specific housing return series. In particular, the equity return series is statistically significant in only 2 of the sampled MSAs; further, the estimated coefficient magnitudes are negligible. Table 3 also provides evidence of substantial cross-sectional variability in model explanatory power and estimated housing return betas, which range upwards to about 75 percent and 2.61, respectively, from a low of near zero. In sum, results of models (1) and (2) suggest the appropriateness of the national housing return series to proxy market returns in the housing pricing model. The findings for model (1) are strongly supportive of the Case-Shiller behavioral studies. Per the asset pricing model, the market variable is an investment variable with its estimated beta coefficient representing the magnitude of market risk. The strong findings in model (1) for the market factor suggest a beta risk and return relationship where investment in housing follows a risk and return strategy; investment in high risk areas is compensated by high returns, whereas investment in low risk areas results in a low return reward. Further, as evidenced in Case-Shiller survey results, there exists substantial variation in housing investment behavior among buyers in different metropolitan specific markets as identified by variability in the estimated market betas.

---

[19] As is common to the empirical asset pricing literature, we also estimate the housing asset pricing models in an excess return specification, whereby the MSA and national house price return series are adjusted by the risk-free rate. In that specification, we use the 3-month Treasury Bill to proxy the risk-free rate. Research findings are robust to the excess return transformation of the model and are not presented for conciseness. Those results are available on request.

[20] The distinct weighting structure of the candidate market factors may have consequences for the inferences of the single factor asset pricing model results. However, given the very strong support for the OFHEO series over the equity series as the market proxy we do not comment on this issue further.

Subsequent models augment the single factor market return specification so as to determine whether there are other risk factors that are compensated by additional returns. In model (3), we estimate a two-factor model which controls for size effects associated with differences in returns between low- and high-priced metropolitan housing markets. Here we test the hypothesis that lower-priced MSA housing markets offer higher risk-adjusted returns than higher-priced MSAs. This term bears a relation to the small firm effect evidenced in the equity pricing literature, whereby small firms offer higher risk-adjusted returns than large firms. This effect is sufficiently prominent so as to be included in standard asset pricing models such as the Fama and French (1993) 3-factor model as a Small minus Big (SMB) variable where the returns from large capitalization stocks are subtracted from those of small stocks and the resulting zero-investment variable is included as an explanatory variable. As suggested above, the housing small minus big term (SMB) is defined as the quarterly return associated with the 25$^{th}$ percentile MSA house price area less that associated with the 75$^{th}$ percentile MSA house price area.[21] Results from Table 3 indicate that the coefficient on the housing small minus big term is precisely estimated only in 19 of the 151 MSAs and is not of the anticipated sign. Accordingly, systematic arbitrage of returns among high- and low-priced MSAs does not appear relevant to housing asset pricing in the vast majority of sampled areas. Note, however, that the estimated market beta is robust to inclusion of this term and the explanatory power of the single factor model increases with its inclusion.[22]

In model (4), we estimate another two-factor model that tests for momentum effects. Consistent with the finance literature (see, for example, Jegadeesh and Titman, 1993), our momentum term is defined as the difference in average house price returns between the lagged 10 highest and 10 lowest return MSAs for each quarter. In the finance literature, this variable has been used to proxy the investment strategy of going long with the previous period's winners while at the same time shorting losers from the prior period in a zero-investment positive return approach. In the housing application, this formulation tests the hypothesis that investors would identify the best performing MSAs in the country and fund investments in those areas via sales of property in the worst performing areas. Accordingly, the momentum term seeks to identify speculative spatial strategies among housing investors. Indeed, this formulation of the momentum term derives as well from Case-Shiller survey findings, which indicate higher (lower) levels of speculative home purchase in rising (falling) housing markets. As evidenced in Table 3, the estimated momentum terms are quite small in magnitude, with an average of slightly less than zero as against a positive prediction, and precisely estimated only in 18 MSAs.[23] Results of the housing investment risk-return framework accordingly do not provide much support for a geographic arbitrage zero-investment strategy, although including momentum does increase the explanatory power of the housing investment model without impacting the influence of the market beta.

[21] We tested this with alternative housing formulations of the SMB and the results were qualitative similar and are available on request.

[22] Note it is possible that the SMB term provides explanatory power in analysis of housing market returns but that its impact is reduced due to the use of MSA-level time-series rather than property-specific data.

In model (5), we estimate a two-factor model which incorporates idiosyncratic risk. As is broadly appreciated, household investment in housing is typically among a small number of properties and is highly undiversified.[24] Liquidity constraints and difficulty in shorting the housing asset further constrain diversification. Accordingly, investment in housing diverges markedly from the usual scenario for equity pricing, where market participants are able to invest in a diversified equity portfolio.[25] The unique aspects of investment in housing suggest that our housing asset pricing model compensate investors for total risk, inclusive of both systematic (market) risk and unsystematic (idiosyncratic) risk. Accordingly, our specification follows that of Merton's (1987) model, where both market risk and idiosyncratic risk require risk-return compensation. Our methodology for computing idiosyncratic risk is standard to the asset pricing literature. We use the standard deviation of squared residuals associated with estimation of the simple single market factor model for each MSA to proxy for MSA-specific non-market returns. As indicated in Table 3, we find weak evidence in support of idiosyncratic risk explaining the variation in house price returns. Here, the idiosyncratic risk proxy enters the asset pricing model with a high level of statistical significance only in 25 of the 151 sampled MSAs. Further, the estimated betas on the market factor appear robust to the inclusion of this term and are very similar to those reported for model (1). The inclusion of idiosyncratic risk does however increase the explanatory power of the model.

Finally, in model (6), we estimate a four factor model that controls for the market factor, idiosyncratic risk, momentum, and size effects. The inclusion of these 4 factors aims to replicate the augmentation of the multi-factor models in equity pricing, namely additional factors have been found to generate anomalous pricing behaviour (see Fama and French, 1996), and as a consequence have been incorporated into the pricing model as potential risk factors. As evidenced in Table 3 and discussed above, estimation results suggest a very strong relation between MSA specific house price returns and market risk. In addition, there is only limited significance for the idiosyncratic risk, momentum, and size terms in the determination of MSA housing returns. Those controls sometimes enter the estimating models with an unanticipated sign; further, they are significant only for a small number of MSAs. Idiosyncratic risk, for example, is significant only in about 15 MSAs. Note, however, that the inclusion of those terms boosts the average explanatory power of the housing asset pricing model with the average explanatory rising to almost 30 percent compared to 20 percent for the single factor model. Notably, the average estimated market beta remains robust to the inclusion of those controls and actually increasing in significance from 103 to 109 MSAs.

[23] Our findings are qualitatively robust to the inclusion of variations in the specified momentum term and are available on request.

[24] The most common for form of housing investment is in a single unit of owner-occupied property.

[25] Although investors in equities do not necessarily exploit full diversification in their investment strategy (see Merton, 1987; and Malkiel and Xu, 2006).

Figure 2 provides a further indication of variation in estimated market betas across sampled MSAs. Plotted in Figure 2 are betas sorted by magnitude from lowest to highest MSAs. To illustrate the variation in magnitude of betas across MSAs, we plot every 10$^{th}$ market beta in the sample. The betas are generated from regression model (1) in Table 3.[26] Also plotted are the 95 percent confidence bands. The housing market betas indicate substantial cross-sectional variation, ranging from -.185 in Provo, UT to a positive 2.61 in Modesto, California. In the case of Modesto, the estimated beta suggests a highly volatile market that moves by 2.61 percent for every percentage point move in the national house price series. Among U.S. metropolitan areas, the California central valley boom town of Modesto recorded the greatest house price response to movements in the National OFHEO series.

The complete set of estimated market betas by MSA is contained in Appendix Table 1. As would be anticipated, elevated betas are estimated for metropolitan areas on the west coast and Florida. Note also that high betas are evidenced across a wide-ranging set of geographical areas in California that are both supply and non-supply constrained. Many metropolitan areas in the Midwest are characterized by low housing market betas. Table 1 also provides information on average housing market returns for each sample MSA. As evidenced in the Table, the statistical significance of the estimated beta and the overall explanatory power of the simple housing asset pricing model tend to be highest in those MSAs with the larger market betas. The large estimated beta associated with Washington, D.C. is highly significant; further, national housing market returns alone explain 75 percent of the variations in Washington, D.C. housing market returns. In marked contrast, the estimated market betas for low beta areas are largely insignificant; typically, the explanatory power of the housing model is quite low in those areas.

Several important conclusions emerge from the MSA-specific results. Firstly, on average, the single factor housing model works well to capture the common variation in MSA housing returns. Results accordingly indicate the relevance of the housing investment framework in an explanation of sampled MSA housing returns. However, as would be expected, the investment asset pricing model is not similarly relevant in all places, as findings indicate a strong geographic dispersion in the magnitude of the estimated market betas and the model fit. For example, consistent with Case-Shiller behavioral survey results, investment considerations, as captured in the housing model, are highly important to the determination of house price returns in many US coastal markets. On the other hand, the investment model has little explanatory power in many smaller, mid-western cities.

We now assess the robustness of the investment model results to the inclusion of controls for local economic variables. Table 4 provides results of an augmented housing asset pricing model that includes explanatory variables commonly cited in the housing literature (see, for example, Case and Shiller (1988, 1990), Goodman and Gabriel (1996), Case and Quigley

[26] Similar variation for market betas occur for the augmented models and are available upon request.

(1991), Gabriel, Mattey and Wascher (1999), and Himmelberg, Mayer, and Sinai (2005)). Those variables (defined above) include controls for quarterly changes in MSA employment, log of lagged nominal housing affordability (defined per convention in terms of nominal house price/income ratios), and change in housing foreclosures. Per above, models (1) – (3) add those terms sequentially, starting with the employment growth proxy, to the four-factor housing framework. As evident in model (1), the employment growth term is significant only in a few metropolitan areas. A similar outcome is evidenced in model (3) for the change in foreclosures term. As is broadly appreciated, nominal affordability is an important input to mortgage qualification and to housing demand. Results of the augmented models indicate that nominal affordability is significant in explanation of house price changes in approximately 50 of the 151 metro areas in our sample. Further, inclusion of the affordability term adds to the explanatory power of the model overall. That notwithstanding, results of the augmented models indicate substantial robustness to the basic single market factor model relationship in the determination of housing returns. Indeed, the estimated market betas remain robust and are significant in some 115 of the 155 MSAs. Overall, consistent with survey research findings, results of the augmented model provide strong support for the housing investment model, where expected returns to housing are related to market risk and where other fundamentals are of secondary importance.

We now turn to questions of temporal variability in the asset pricing model results. Figures 3 – 6 plot the temporal variation in market betas and model $R^2$ for San Francisco, Boston, Milwaukee, and Los Angeles. The respective plots also illustrate the cross sectional variation among the four areas in the boom and bust cycle of US housing. Recall per above that the four markets are the metropolitan areas of focus in the Case-Shiller behavioral research. The estimates displayed in these figures derive from a 24 quarter moving estimation period.[27] Further, the model (1) housing framework is used to compute the estimates. For the most part, the results are suggestive of the market factor having substantial explanatory power in those cities throughout the sample, with the exception of Milwaukee in the early part of the timeframe. In all cases, both the estimated market betas and the model explanatory power change dramatically over time. The plots are further instructive in discerning the relationship between the magnitude of the estimated market factor and the model explanatory power. The investment model works best for those MSAs associated with the highest level of speculative behavior and for periods with the highest levels of speculative activity. In that regard, as noted in memo items to the plots, the simple correlations between $R^2$ and the estimated market betas range from about 0.616 in San Francisco to 0.961 in Milwaukee. Indeed, both the estimated betas and the model explanatory power spike during periods of housing market boom. In San Francisco, the estimated market betas increase past 2, with model fit of about 80 percent, during the most recent housing boom of the late 2000s.

However, as housing booms turn to economic downturn and housing bust, both the investment model explanatory power and estimated market betas fall markedly. The extreme cyclical variability in these terms is also evidenced in Boston. Estimated market betas and $R^2$ trended down markedly during the first half of the current decade—suggesting markedly diminished importance of the risk-return characterization to Boston house price fluctuations during that period. While those trends reversed in the context of the recent housing boom, estimated market betas in Boston failed to reach levels recorded for coastal California. Finally, Milwaukee presents a different case altogether. Consistent with early Case-Shiller behavioral findings (Case and Shiller, 1988), the risk-return housing investment model, as embodied in the housing asset pricing model, provides little insight as to Milwaukee house price trends for much of the 1990s. Indeed, in early years, both estimated market betas and model explanatory power approximated zero. For the decade of the 1990s, as described by Case and Shiller, one would be hard-pressed to argue the importance of investment demand as deterministic for housing market fluctuations in Milwaukee. Interestingly, as evidenced by findings from the most recent housing boom, both investment model explanatory power and magnitude of estimated market beta jumped sharply for Milwaukee. Consistent with 2009 Case-Shiller survey findings, results from Milwaukee typify the substantially broader applicability of the housing risk-return framework in explanation of housing market fluctuations in recent years.

[27] We also experimented with variation of the timeframe of the moving estimation period from 16- to 30-quarters. Note that results are largely robust to those specification changes.

### 2.4 Fama-MacBeth

The Fama-MacBeth (1973) framework has been extensively applied to test the validity and related implications of the asset pricing models. Specifically, the Fama-MacBeth approach allows us to examine a number of distinct implications for our asset pricing model. Of particular importance is assessment of whether there exists a significant positive market beta-return relationship, implying that the market beta can explain the variation in MSA housing returns and that the variation is positively related to beta. Accordingly, the Fama-McBeth framework tests whether beta risk is the important driver of MSA housing returns. The Fama-MacBeth approach also allows us to test whether idiosyncratic risk, as proxied by the standard deviation of the single market factor model's squared residuals, is related to asset returns.[28] Further, the Fama-MacBeth framework allows us to determine whether there are non linearities in the beta-return relationship.

Similar to much of the asset pricing testing in equity markets, the Fama-MacBeth framework uses portfolios. As is broadly appreciated, the use of portfolios helps to avoid problems of errors in variables (see Miller and Merton, 1972). That notwithstanding, the use of portfolios involves choices regarding portfolio composition which can influence the outcome of the analysis (see Brennan, Chordia and Subrahmanyam, 1996). Moreover, the use of portfolios can aggregate valuable information about the individual assets that comprise the portfolio (Roll, 1977). Accordingly, the Fama-MacBeth approach is not without its faults. Some of the challenge is due to estimation of parameters (e.g. see Shanken, 1992), whereas a more fundamental concern is associated with identification of the market portfolio and the use of related proxies (see Roll, 1977; and Roll and Ross, 1984). Notwithstanding these concerns, the Fama-MacBeth framework is the standard approach to testing the validity of the single-factor and related multi-factor models (Brennan, Chordia and Subrahmanyam, 1998).

To motivate the asset pricing tests Figure 7 presents the full set of MSA market betas and their associated mean returns using model (1) in Table 3. The figure provides graphic evidence of a positive market risk and return relationship with high beta MSAs associated with high return MSAs. For the indentified MSAs in the scatter plot such as Salinas we see a high beta and high average returns whereas in contrast for Dallas-Plano-Irving we see a low beta and low average returns. However, as evidenced in the chart, there is not a very precise positive linear beta return relationship with some MSAs having a low beta and high returns. The lack of goodness of fit of the universe of MSA returns has a number of possible interpretations. First, it may imply that the OFHEO equally weighted index is not mean-variance efficient (Shanken, 1985; Gibbons, Ross and Shanken, 1989). Along the same vein however, the scatter may be due to small biases in the parameter estimates where the true values would result in a linear relationship between the market portfolio and the set of MSA returns (Levy and Roll, 2009). Alternatively, the plot may be indicative of a role for non-market risk that results in deviations from a positive linear beta return relationship. In the Fama-MacBeth approach, non-market risk is assumed to be unidentified idiosyncratic risk.

[28] For the asset pricing model to hold in its strictest sense of having a single market factor explaining house price returns, one would anticipate that non-market risk would be an insignificant determinant of MSA returns in the housing application, however, few investors likely hold a diversified portfolio. Hence, we hypothesize that housing investors require a return for both market and idiosyncratic risk.

Following the Fama-MacBeth approach we identify 3 steps. First, in quarters 1-30 we identify the market betas for each MSA implied by model (1) in Table 3. These individual betas are sorted by rank into 15 portfolios in each period to minimize the errors in variable problem associated with using individual asset betas. Second, post-ranking portfolio betas are estimated in non-overlapping quarters 31-60 using a simple single factor model on the constructed portfolios. Finally, we run cross sectional regressions of the full Fama-McBeth specification for quarters 61-92. That final analysis yields estimated parameters for these quarters that are then used to test the implications of the housing asset pricing model. The specification of the Fama-McBeth portfolio model is indicated in the memo below Table 5.[29]

The Fama-MacBeth results for US housing data are presented in Table 5. Overall, estimates of the model provide strong evidence in support of a positive beta-return relationship in US housing markets. We examine the robustness of these results using data for the full time frame as well as for sub-periods after 2000. In all samples, with the exception of March 2005 and June 2007 sub-period, we obtain strong evidence in support of a positive risk and return relationship for US house prices with an average $\gamma_1$ parameter significantly greater than zero. This relationship is particularly strong between 2002 and 2005. However, our evidence on other Fama-Macbeth tests is not as conclusive. Unlike the premise of a linear risk and return relationship in much of asset pricing, we find strong evidence of a non-linear relationship between returns and beta although there is ambiguity regarding the sign on the estimated $\gamma_2$ parameter. Finally, the Fama-McBeth analysis also provides evidence in support of the non-market (idiosyncratic) risk factor, which we proxied using the standard deviation of squared residuals from model (1) in Table 3. In that regard, the role of non-market risk may thus be masked in our earlier tests, where we present a single set of results for all individual MSAs. Overall however, the primary premise of our asset pricing model is strongly supported in application to housing investment, risk, and return.

### 2.5 Robustness Check

We test the robustness of the model findings by using Case Shiller city specific data. Due to data limitations, we conduct this analysis on the 16 published C-S city specific indexes from 1990 as dependent variables. We use the same models outlined in Tables 3 and 4. For the explanatory variables, we match the MSA level data to the associated Case-Shiller city house price data. Results for the Case-Shiller data are given in Table 6 provide strong support for the investment model where the market factor is the primary determinant of house price returns. We identify the appropriate market return by comparing the Case-Shiller National Index with the SP500 and again find that the stock market is not the relevant benchmark for housing. The Case-Shiller National Index is significant for all but 1 city index and on average has high explanatory power, whereas in contrast the impact of the SP500 on metro area house price indexes is negligible.

[29] We examined various combinations of formation, estimation and testing periods. The results are qualitatively the same and are available on request.

In models (3) through (6) we add further investment related factors and determine their influence on city specific index returns. Similar to our modelling of OFHEO data, the addition of investor momentum, size effects and idiosyncratic risk have only limited influence on the city return data. Although the SMB is statistically significant in 7 of the 16 indexes when examined alone, its influence deteriorates when further variables are added. In contrast, the market factor remains influential across the spectrum of cities. The additional controls do increase the power of the asset pricing model. By including the three extra variables common cited in the literature, the average explanatory power of the model increases to about 60 percent.

Our final set of models, (7) – (9), incorporate socio-economic variables, change in employment, affordability and change in foreclosures as potential explanatory factors for the Case-Shiller city index data. We find weak evidence for these variables in explaining the expected returns of city level data. Specifically, employment growth is influential for 5 cities whereas change in affordability is statistically significant for 4 indexes. Again, inclusion of these terms serves to raise the explanatory power of the housing asset pricing model to almost 67 percent on average. Moreover, the market beta remains largely robust and is clearly the dominating explanatory variable in the pricing model. Overall testing with the Case-Shiller city data confirms the findings of our asset pricing model as applied to the OFHEO series.

## 3. Conclusions

In this research, we assess the importance of the risk-return framework in determination of metropolitan housing returns. To do so, we apply quality-adjusted house price data from 151 U.S. metropolitan areas over the 1985-2007 period to estimate a multi-factor housing asset pricing model. Overall, results indicate a sizable and statistically significant influence of the market factor on MSA house price returns. Further, the single factor housing model is largely robust to the addition of other explanatory terms, including measures of idiosyncratic risk, momentum, geographic arbitrage among high- and low-priced metropolitan areas, and other housing market fundamentals. Our market betas vary substantially and are strongly related to the relative explanatory power of the models in the cross-section. Further, our results suggest considerable time-variation in housing model explanatory power, with markedly elevated importance of the pricing framework over the course of the recent house price cycle. Results are largely robust to the use of OFHEO MSA specific and Case-Shiller city-specific data.

To avoid potential errors-in-variables problems associated with the use of single assets, we apply the Fama-MacBeth framework to examine the pricing relationship among portfolios of MSA returns. We again find a strong positive risk and return relationship for the portfolios. However, results of that analysis suggest some non-linearity to the risk-return relationship as well as indicate the importance of idiosyncratic risk to housing asset pricing. In marked contrast to reduced form specifications which largely focus on consumption aspects of housing demand, our findings are supportive of the application of a housing investment risk-return model in explanation of variation in metro-area cross-section and time-series of US house price returns. The findings strongly corroborate Case-Shiller behavioral findings indicating the importance of speculative forces in the determination of U.S. housing returns.

**Figure 1**
**National and Individual MSA House Price Indices**

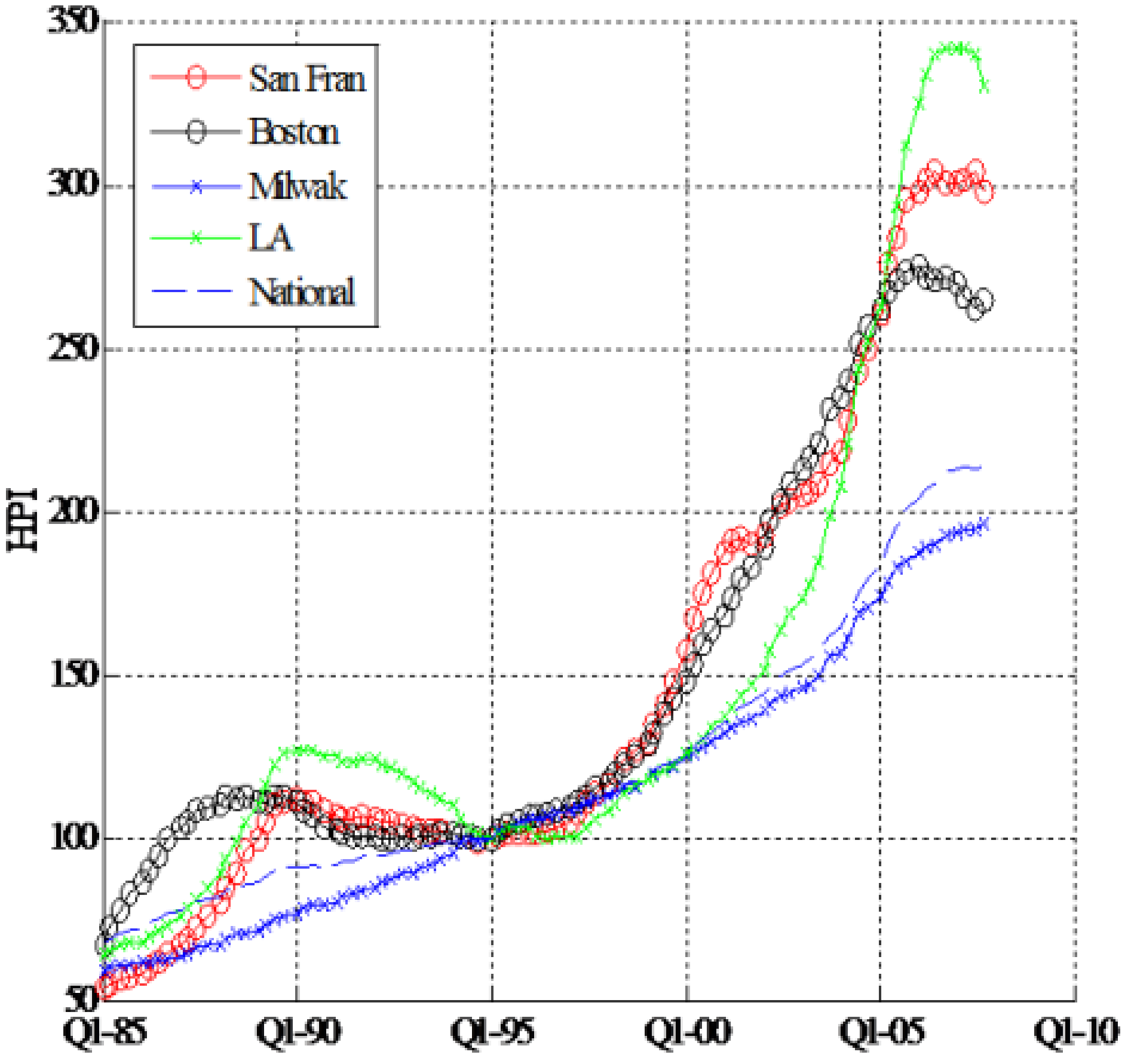

**$R_{HPI}$ Summary Statistics by Select MSA**

| | Mean | Std Dev | Min | Max |
|---|---|---|---|---|
| **San Francisco** | 1.883 | 2.217 | -2.378 | 6.691 |
| **Boston** | 1.503 | 2.071 | -3.274 | 7.114 |
| **Milwaukee** | 1.31 | 0.801 | -0.333 | 4.707 |
| **Los Angeles** | 1.798 | 2.549 | -4.333 | 9.953 |
| **National** | 1.257 | 0.781 | -0.391 | 3.742 |

The plot details the time series of quarterly index levels for 4 individual MSAs and for the National OFHEO series between 1985 and 2007. The table contains quarterly summary statistics of percentage returns ($R_{HPI}$) for 4 individual MSAs and for the National OFHEO series between 1985 and 2007.

**Figure 2**
**Plot of Select Market Betas**

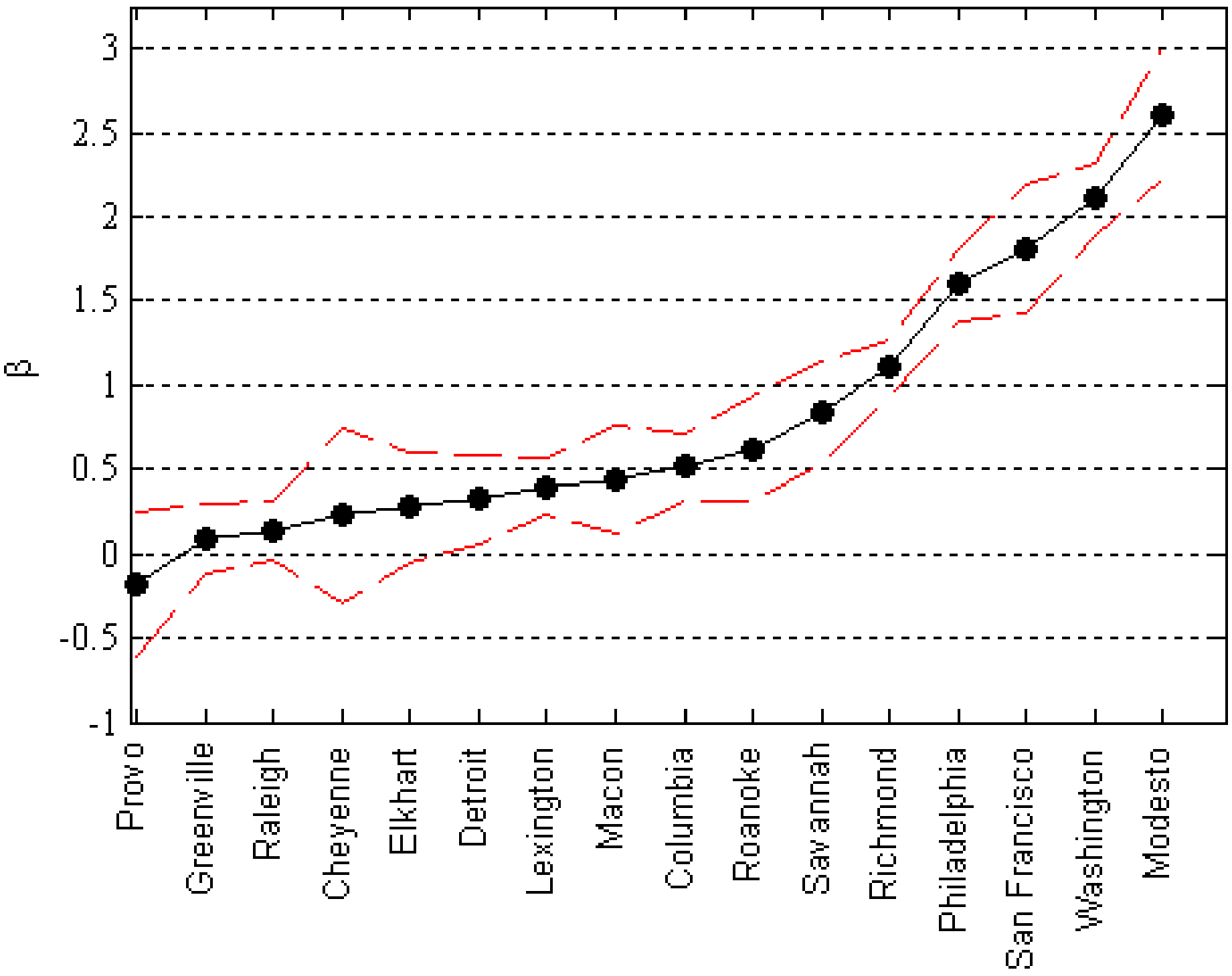

The plot is based on a sort of betas by magnitude from lowest to highest and where every 10th beta is selected for presentation. The betas are obtained from model (1) in Table 3. The 95 percent confidence bands of the identified MSA betas are also given (dashed lines).

**Figure 3**
**Temporal Variation in Market Betas and Model Explanatory Power**
**San Francisco**

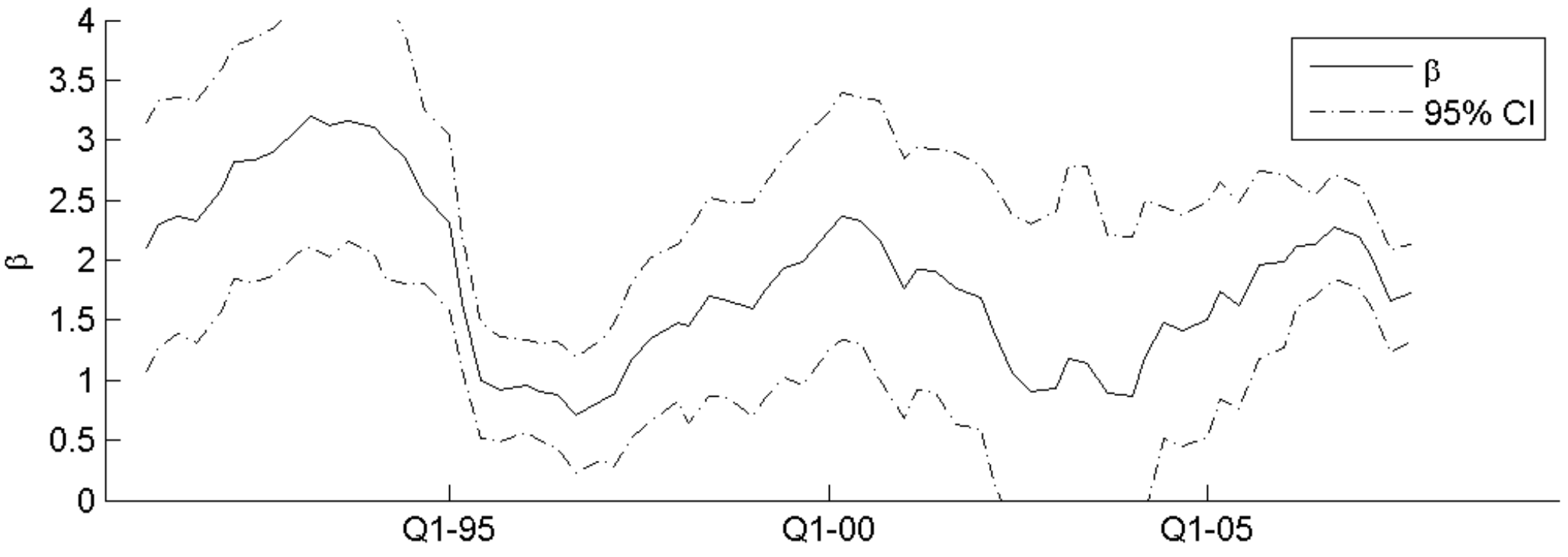

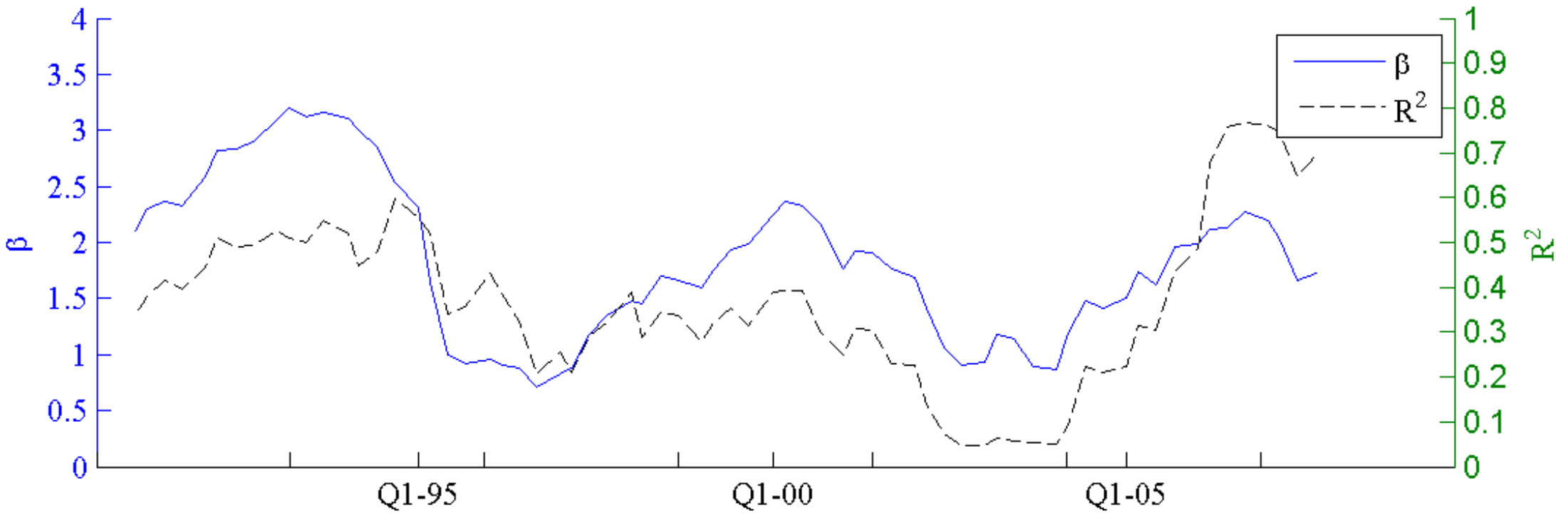

| | Mean | Std Dev | Min | Max |
|---|---|---|---|---|
| **ß** | 1.824 | 0.692 | 0.663 | 3.209 |
| **$R^2$** | 0.373 | 0.192 | 0.037 | 0.796 |
| **Corr(ß,$R^2$)** | 0.616 | | | |

The plots detail the time series of quarterly market betas for San Francisco using model (1) in Table 3. The top plot provides a 95% confidence band on the estimated market betas. The bottom plot includes the market betas and associated $R^2$. The timeframe is between 1985 and 2007 based on a 24 quarter moving window and where the initial betas are obtained for 1991. The table contains quarterly summary statistics of the associated market betas and model $R^2$ for San Francisco using model (1) in Table 3. The correlation between market betas and $R^2$ is also given.

**Figure 4**
**Temporal Variation in Market Betas and Model Explanatory Power**
**Boston**

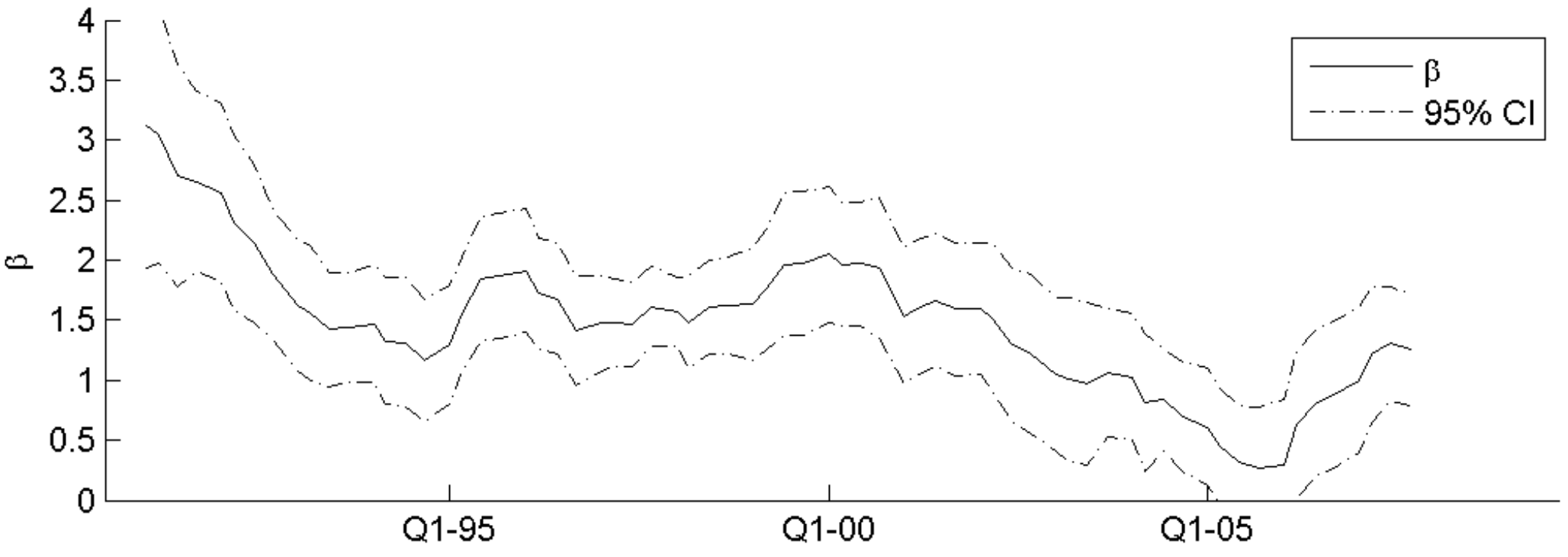

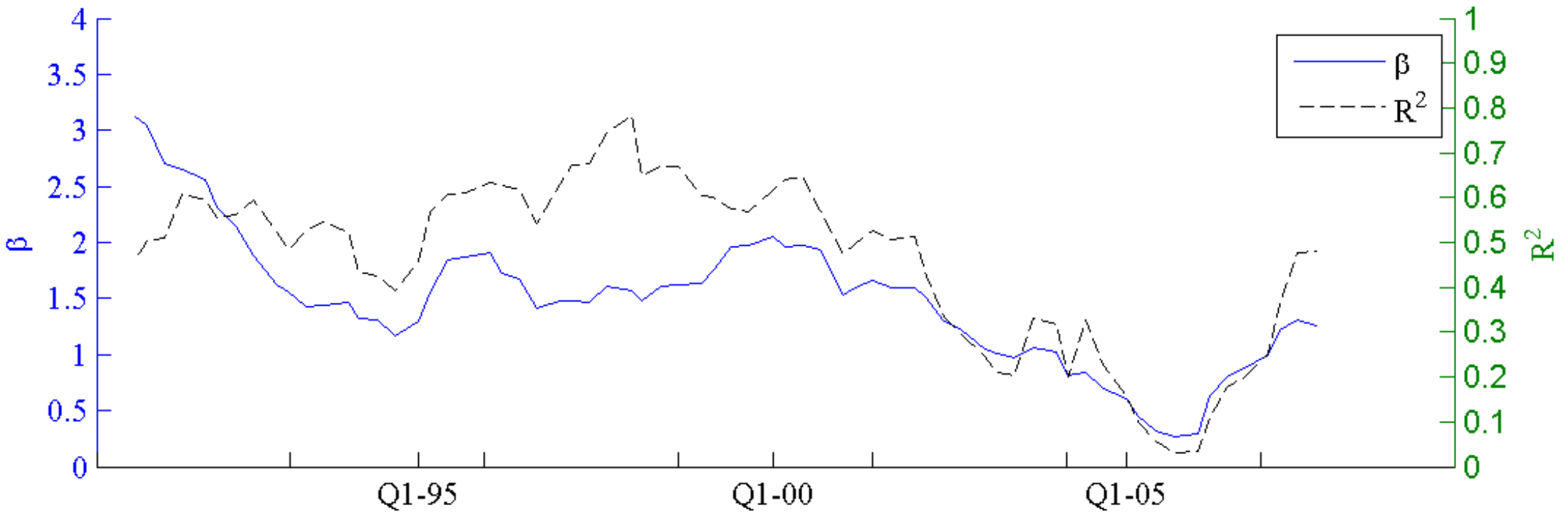

| | Mean | Std Dev | Min | Max |
|---|---|---|---|---|
| **ß** | 1.479 | 0.586 | 0.269 | 3.061 |
| **$R^2$** | 0.461 | 0.187 | 0.035 | 0.785 |
| **Corr(ß,$R^2$)** | 0.704 | | | |

The plots detail the time series of quarterly market betas for Boston using model (1) in Table 3. The top plot provides a 95% confidence band on the estimated market betas. The bottom plot includes the market betas and associated $R^2$. The timeframe is between 1985 and 2007 based on a 24 quarter moving window and where the initial betas are obtained for 1991. The table contains quarterly summary statistics of the associated market betas and model $R^2$ for Boston using model (1) in Table 3. The correlation between market betas and $R^2$ is also given.

**Figure 5**
**Temporal Variation in Market Betas and Model Explanatory Power**
**Milwaukee**

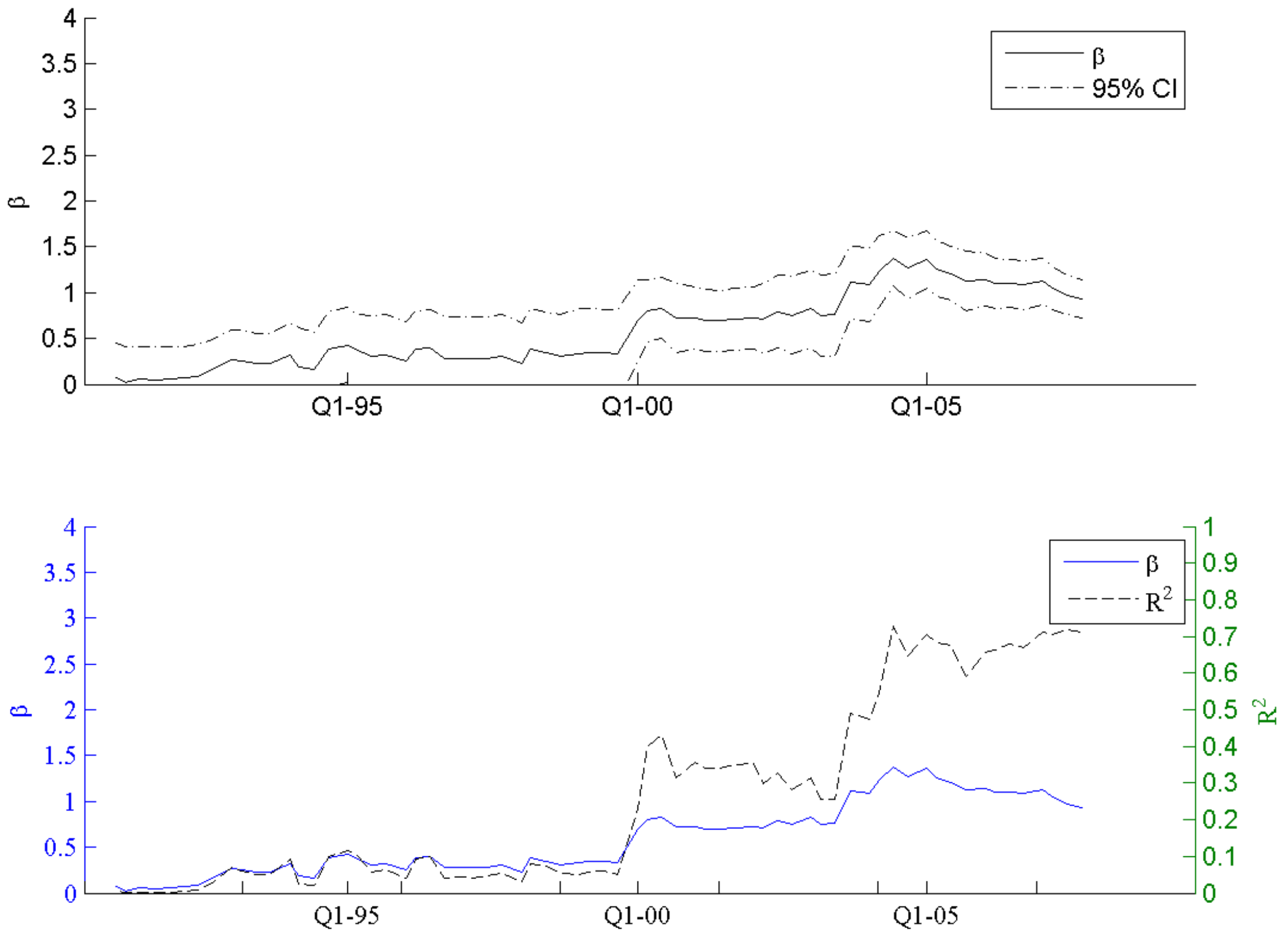

| | Mean | Std Dev | Min | Max |
|---|---|---|---|---|
| **β** | 0.594 | 0.396 | 0.048 | 1.391 |
| **$R^2$** | 0.262 | 0.258 | 0.002 | 0.719 |
| **Corr(β,R** | 0.961 | | | |

The plots detail the time series of quarterly market betas for Milwaukee using model (1) in Table 3. The top plot provides a 95% confidence band on the estimated market betas. The bottom plot includes the market betas and associated $R^2$. The timeframe is between 1985 and 2007 based on a 24 quarter moving window and where the initial betas are obtained for 1991. The table contains quarterly summary statistics of the associated market betas and model $R^2$ for Milwaukee using model (1) in Table 3. The correlation between market betas and $R^2$ is also given.

**Figure 6**
**Temporal Variation in Market Betas and Model Explanatory Power**
**Los Angeles**

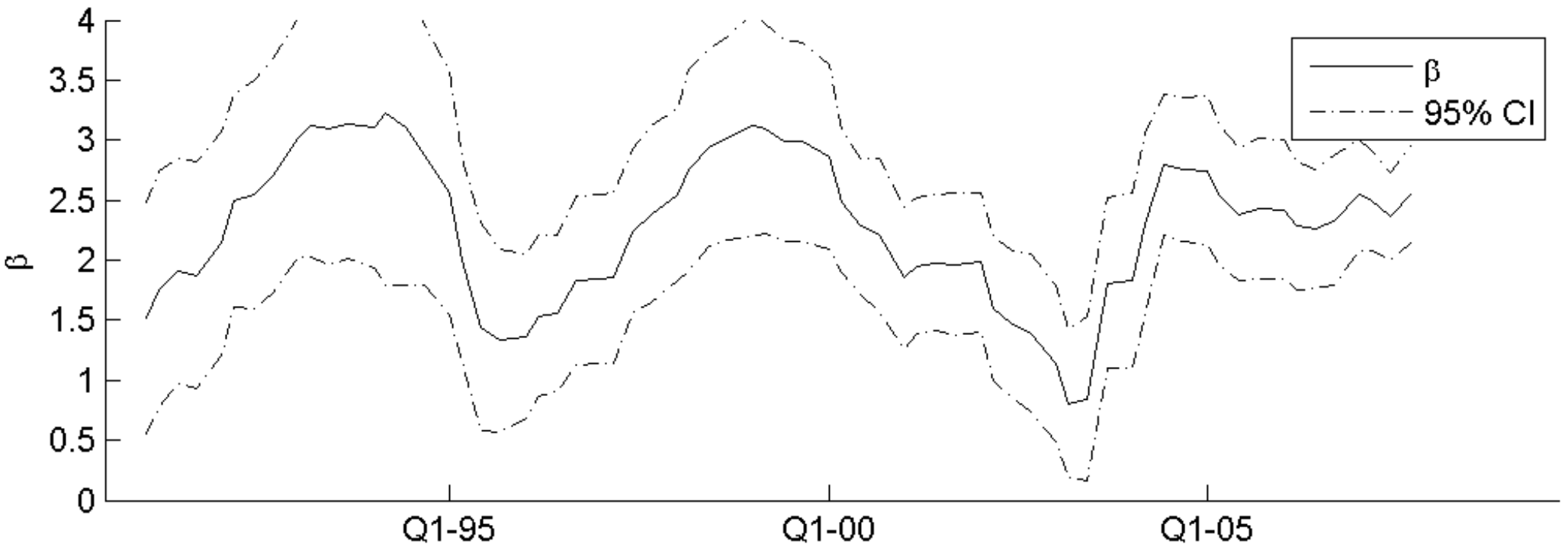

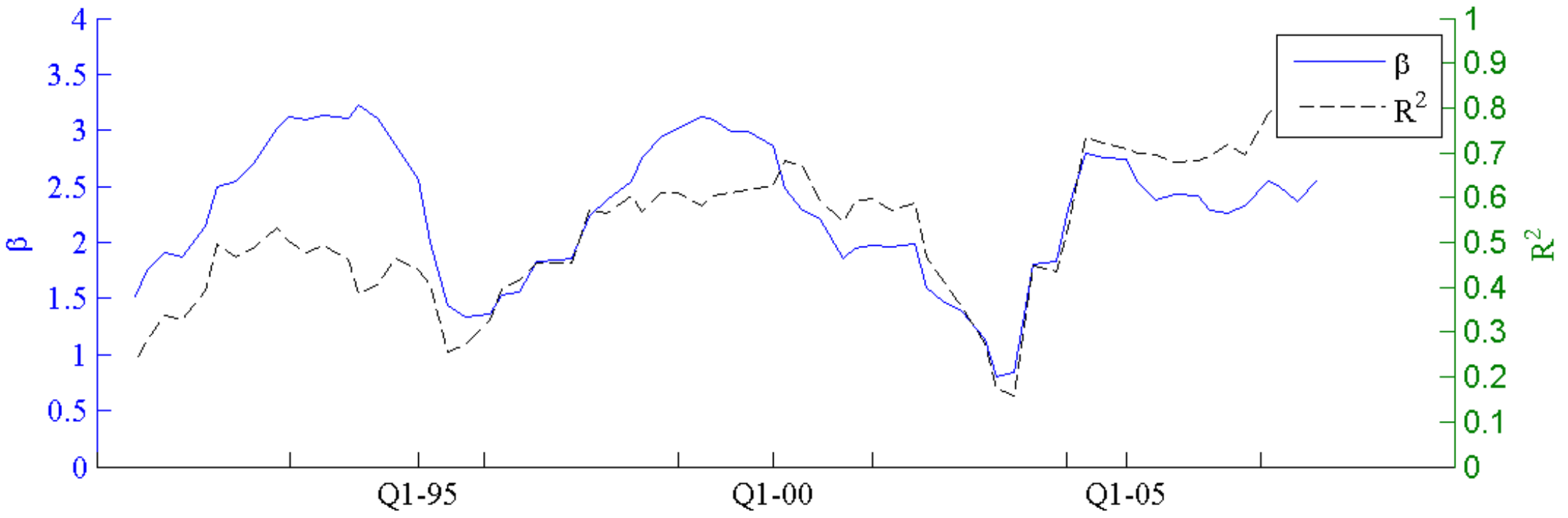

| | Mean | Std Dev | Min | Max |
|---|---|---|---|---|
| **ß** | 2.232 | 0.627 | 0.807 | 3.14 |
| **$R^2$** | 0.516 | 0.168 | 0.158 | 0.832 |
| **Corr(ß,$R^2$)** | 0.635 | | | |

The plots detail the time series of quarterly market betas for Los Angeles using model (1) in Table 3. The top plot provides a 95% confidence band on the estimated market betas. The bottom plot includes the market betas and associated $R^2$. The timeframe is between 1985 and 2007 based on a 24 quarter moving window and where the initial betas are obtained for 1991. The table contains quarterly summary statistics of the associated market betas and model $R^2$ for Los Angeles using model (1) in Table 3. The correlation between market betas and $R^2$ is also given.

**Figure 7**
**MSA Housing Market Risk and Return**

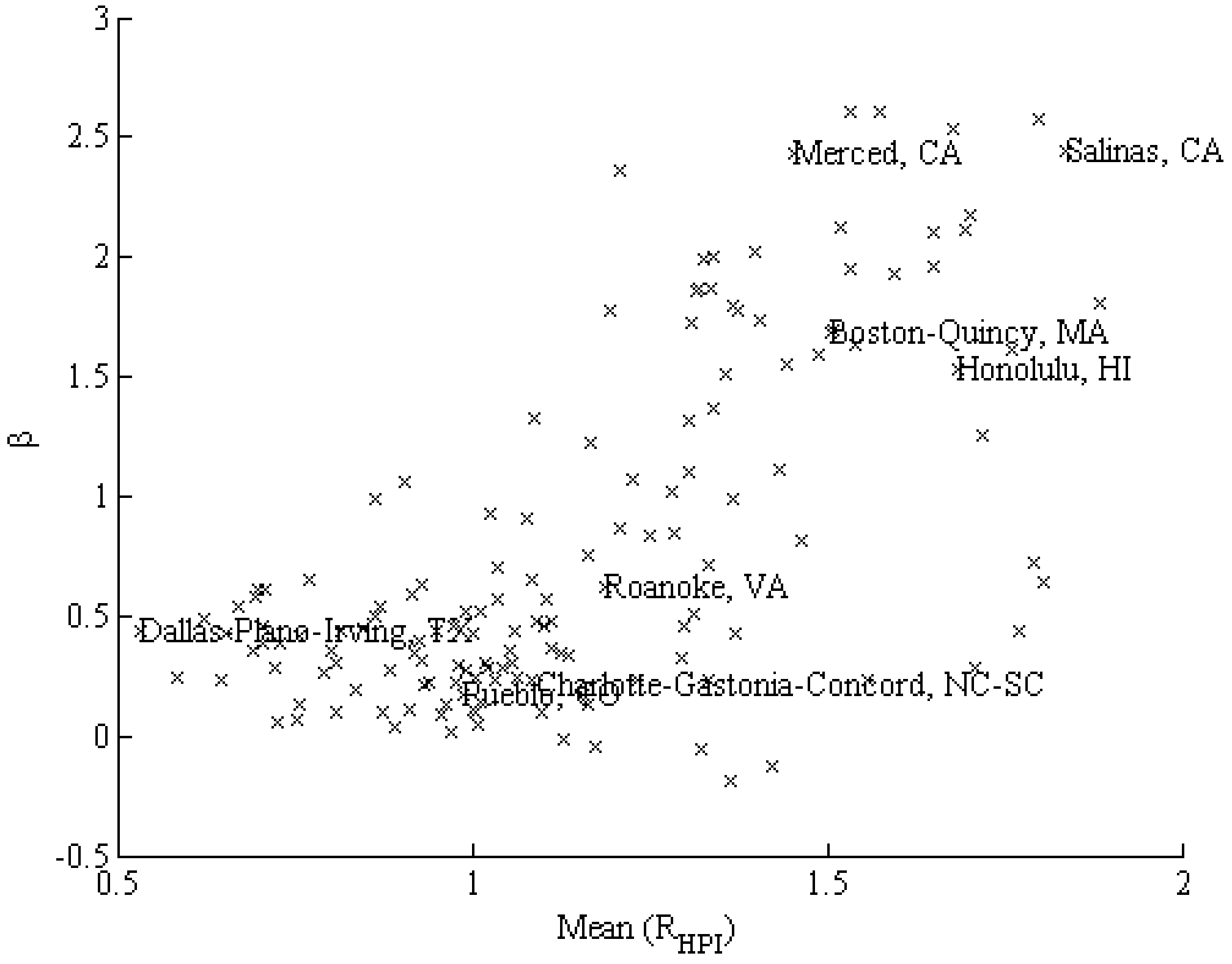

The scatter plot shows the full sample of MSA market betas and their respective mean returns. The timeframe is between 1985 and 2007 where the betas are obtained from model (1) in Table 3 and the mean returns are in Appendix Table 1.

**Table 1**
**Summary Statistics Presented at Yearly Frequency**

| | Mean | SD | Min | Median | Max |
|---|---|---|---|---|---|
| $\mathbf{R_{HPI}}$ | 0.924 | 0.738 | -0.295 | 0.854 | 2.530 |
| $\mathbf{R_{OFHEO}}$ | 1.150 | 0.754 | -0.391 | 1.190 | 2.870 |
| $\mathbf{R_{SP}}$ | 9.67 | 17.5 | -32.2 | 11.3 | 33.0 |
| **SMB** | 0.175 | 0.406 | -0.882 | 0.135 | 0.827 |
| **Mom** | 6.35 | 2.97 | 2.98 | 5.51 | 12.7 |
| $\mathbf{s^2}$ | 4.590 | 4.530 | 0.693 | 3.170 | 19.700 |
| **ΔEmp** | -0.764 | 0.329 | -1.270 | -0.563 | -0.120 |
| **Log(Afford$_{t-1}$)** | 0.241 | 0.011 | 0.216 | 0.244 | 0.257 |
| **ΔForc** | 1.090 | 0.217 | 0.802 | 1.060 | 1.720 |

The variables are defined as follows. $R_{HPI}$ is the asset return for each MSA: 100*[log(HPI$_t$)-log(HPI$_{t-1}$)]. $R_{OFHEO}$ is a market return proxy: 100*[log(NatHPI$_t$)-log(NatHPI$_{t-1}$)]. $R_{SP}$ is a market return proxy: 100*[log(SP500$_t$)-log(SP500$_{t-1}$)]. SMB (Small Minus Big) is the return of 25th percentile house price MSA minus the return of 75th percentile house price MSA. Mom (Momentum) is the one quarter lagged average return of 90th percentile MSAs minus the one quarter lagged average return of 10th percentile MSAs. $s^2$ (Idiosyncratic Risk) is the standard deviation of squared residuals from Model 1 (Table 3). ΔEmp is the change in employment for each MSA: 100*[log(Emp$_t$)-log(Emp$_{t-1}$)]. Afford is the lagged affordability for each MSA: log(Income/Price)$_{t-1}$. ΔForc is the change in foreclosures for each MSA: 100*[log(Forc$_t$)-log(Forc$_{t-1}$)].

Annualized summary statistics are presented for the housing asset pricing model variables with the associated definitions for the timeframe between 1985 and 2007. Time series summary statistics are provided for the variables with no MSA specific data ($R_{OFHEO}$ and $R_{SP}$). Time series summary statistics are presented for the cross-sectional means for the other variables with MSA specific data.

**Table 2**
**Correlation Matrix of Time Varying Terms**

| | $R_{HPI}$ | $R_{OFHEO}$ | $R_{SP}$ | SMB | Mom | $s^2$ | ΔEmp | Afford | Δforc |
|---|---|---|---|---|---|---|---|---|---|
| $R_{HPI}$ | 1.000 | | | | | | | | |
| $R_{OFHEO}$ | 0.953 | 1.000 | | | | | | | |
| $R_{SP}$ | -0.029 | -0.023 | 1.000 | | | | | | |
| SMB | 0.402 | 0.300 | -0.019 | 1.000 | | | | | |
| Mom | 0.287 | 0.234 | 0.103 | 0.660 | 1.000 | | | | |
| $s^2$ | 0.047 | -0.106 | 0.024 | 0.652 | 0.424 | 1.000 | | | |
| ΔEmp | 0.092 | 0.028 | 0.047 | 0.184 | -0.091 | 0.091 | 1.000 | | |
| Afford | -0.499 | -0.412 | 0.063 | -0.509 | -0.522 | -0.179 | 0.038 | 1.000 | |
| ΔForc | 0.431 | 0.512 | -0.210 | 0.038 | 0.108 | -0.135 | -0.040 | -0.352 | 1.000 |

The variables are defined as follows. $R_{HPI}$ is the asset return for each MSA: $100*[\log(HPI_t)-\log(HPI_{t-1})]$. $R_{OFHEO}$ is a market return proxy: $100*[\log(NatHPI_t)-\log(NatHPI_{t-1})]$. $R_{SP}$ is a market return proxy: $100*[\log(SP500_t)-\log(SP500_{t-1})]$. SMB (Small Minus Big) is the return of 25th percentile house price MSA minus the return of 75th percentile house price MSA. Mom (Momentum) is the one quarter lagged average return of 90th percentile MSAs minus the one quarter lagged average return of 10th percentile MSAs. $s^2$ (Idiosyncratic Risk) is the standard deviation of squared residuals from Model 1 (Table 3). ΔEmp is the change in employment for each MSA: $100*[\log(Emp_t)-\log(Emp_{t-1})]$. Afford is the lagged affordability for each MSA: $\log(Income/Price)_{t-1}$. ΔForc is the change in foreclosures for each MSA: $100*[\log(Forc_t)-\log(Forc_{t-1})]$.

A correlation matrix is presented for the housing asset pricing model variables with the associated definitions for the timeframe between 1985 and 2007.

**Table 3**
**Housing Asset Pricing**
**Investment Models**

$R_{HPI} = \alpha_0 + \beta R_m + x'\delta + \varepsilon$

| | (1) | (2) | (3) | (4) | (5) | (6) |
|---|---|---|---|---|---|---|
| $R_{OFHEO}$ | 0.785 | | 0.809 | 0.809 | 0.792 | 0.81 |
| | (103) | | (104) | (105) | (103) | (109) |
| $R_{SP}$ | | -0.001 | | | | |
| | | (2) | | | | |
| **SMB** | | | -0.046 | | | 0.224 |
| | | | (19) | | | (18) |
| **Mom** | | | | -0.003 | | -0.019 |
| | | | | (18) | | (9) |
| $s^2$ | | | | | -0.012 | 0.011 |
| | | | | | (25) | (15) |
| **Distribution of β** | | | | | | |
| **mean** | 0.785 | -0.001 | 0.809 | 0.809 | 0.792 | 0.81 |
| **min** | -0.185 | -0.102 | -0.058 | -0.056 | -0.135 | -0.075 |
| **median** | 0.473 | 0.003 | 0.538 | 0.533 | 0.495 | 0.535 |
| **max** | 2.61 | 0.067 | 2.58 | 2.57 | 2.6 | 2.61 |
| **Distribution of $R^2$** | | | | | | |
| **mean** | 0.191 | 0.007 | 0.24 | 0.243 | 0.236 | 0.277 |
| **min** | 0 | 0 | 0 | 0.001 | 0.005 | 0.012 |
| **median** | 0.098 | 0.003 | 0.161 | 0.164 | 0.156 | 0.208 |
| **max** | 0.752 | 0.055 | 0.799 | 0.796 | 0.783 | 0.81 |

The mean coefficients values for variables of the models listed are presented for the 151 MSAs. The numbers of MSAs from the sample with significant coefficients at the 5% level follow in parentheses. Summary details of the distribution of model betas and the distribution of $R^2$ follow. All models include an unreported constant. The timeframe is between 1985 and 2007 using quarterly data. The variables are defined in Table 1.

**Table 4**
**Housing Asset Pricing**
**Augmented Models**

$$R_{HPI} = \alpha_0 + \beta R_m + x'\delta + z'\gamma + \varepsilon$$

| | (1) | (2) | (3) |
|---|---|---|---|
| $R_{OFHEO}$ | 0.809 | 0.882 | 0.878 |
| | (104) | (112) | (112) |
| SMB | -0.015 | 0.087 | 0.088 |
| | (19) | (19) | (20) |
| Mom | 0.013 | -0.017 | -0.012 |
| | (19) | (20) | (19) |
| $s^2$ | 0.032 | 0.041 | 0.038 |
| | (28) | (31) | (33) |
| $\Delta$Emp | 0.037 | 0.032 | 0.032 |
| | (10) | (7) | (7) |
| Afford | | 2.34 | 2.39 |
| | | (47) | (43) |
| $\Delta$Forc | | | 0.002 |
| | | | (14) |
| **Distribution of $\beta$** | | | |
| mean | 0.809 | 0.882 | 0.878 |
| min | -0.814 | -1.35 | -1.44 |
| %50Q | 0.714 | 0.725 | 0.411 |
| max | 2.4 | 2.45 | 0.723 |
| **Distribution of $R^2$** | | | |
| mean | 0.404 | 0.45 | 0.466 |
| min | 0.0121 | 0.072 | 0.0894 |
| %50Q | 0.345 | 0.382 | 0.393 |
| max | 0.902 | 0.907 | 0.908 |

The mean coefficients values for variables of the models listed are presented for the 151 MSAs. The numbers of MSAs from the sample with significant coefficients at the 5% level follow in parentheses. Summary details of the distribution of model betas and the distribution of $R^2$ follow. All models include an unreported constant. The timeframe is between 1985 and 2007 using quarterly data. The variables are defined in Table 1.

**Table 5**
**Fama-Macbeth Housing Asset Pricing Model Validity Test Statistics**

| **Period** | | **Mean($\gamma_1$)** | **Mean($\gamma_2$)** | **Mean($\gamma_3$)** |
|---|---|---|---|---|
| **Full Period** | | 0.906 | -1.180 | 2.180 |
| | | (2.100) | ( -2.530) | (4.750) |
| **Begin** | **End** | | | |
| **March-00** | **June-02** | 0.254 | -1.190 | 2.410 |
| | | (2.790) | (-3.430) | (12.900) |
| **Sep-02** | **Jan-05** | 1.910 | -2.270 | 3.020 |
| | | (3.830) | (-4.470) | (6.330) |
| **Mar-05** | **Jun-07** | 0.558 | -0.077 | 1.110 |
| | | (1.170) | (-0.1540) | (1.500) |

The included gammas are the average of the estimated gammas in each period from the following model: $R_{portfolio,i,t} = \gamma_{1t}\beta_{it}+\gamma_{2t}\beta_{it}^2+\gamma_{3t}s_t^2+u_t$ . The time-series averages of the estimated gammas are presented and the t-statistics of those averages follow in parentheses. In periods 1-30, we estimate betas for each MSA. Those estimated betas are sorted into 15 portfolios of 10 MSAs each. Using the sorted data, 30 time-series regressions are run for each portfolio based on Model 1 of Table 3. Using the time-series of 30 betas from those regressions, 30 cross sectional regressions are estimated in the testing period, which consists of quarters 61-92.

**Table 6**
**Housing Asset Pricing**
**Case-Shiller City Investment Models**

$$R_{HPI} = \alpha_0 + \beta R_m + x'\delta + \varepsilon$$

| | (1) | (2) | (3) | (4) | (5) | (6) |
|---|---|---|---|---|---|---|
| **$R_{CS\ National\ 16}$** | 0.722 | | 0.641 | 0.681 | 0.618 | 0.622 |
| | (15) | | (13) | (15) | (15) | (13) |
| **$R_{SP}$** | | -0.041 | | | | |
| | | (0) | | | | |
| **SMB** | | | 0.395 | | | -0.0178 |
| | | | (7) | | | (3) |
| **Momentum** | | | | 0.083 | | 0.02 |
| | | | | (3) | | (1) |
| **$s^2$** | | | | | 0.061 | 0.058 |
| | | | | | (6) | (6) |
| **Distribution of CAPM β** | | | | | | |
| **mean** | 0.722 | -0.0412 | 0.641 | 0.681 | 0.618 | 0.622 |
| **min** | 0.119 | -0.095 | -0.068 | 0.096 | -0.148 | -0.136 |
| **%50Q** | 0.837 | -0.034 | 0.511 | 0.723 | 0.499 | 0.476 |
| **max** | 1.34 | 0 | 1.35 | 1.37 | 1.33 | 1.45 |
| **Distribution of $R^2$** | | | | | | |
| **mean** | 0.503 | 0.032 | 0.534 | 0.516 | 0.554 | 0.591 |
| **min** | 0.0421 | 0 | 0.103 | 0.097 | 0.109 | 0.166 |
| **%50Q** | 0.555 | 0.026 | 0.571 | 0.543 | 0.593 | 0.616 |
| **max** | 0.882 | 0.109 | 0.891 | 0.885 | 0.886 | 0.892 |

The mean coefficient values for variables of the models listed are presented for the 16 Case-Shiller cities. The number of cities from the sample with significant coefficients at the 5% level follows in parentheses. Summary details of the distribution of model betas and the distribution of $R^2$ follow. All models include an unreported constant. The timeframe is between 1990 and 2007 using quarterly data. The variables are defined as in Table 1 except for $R_{HPI}$ which here is the Case Shiller housing return series for each city, and $R_{CS\ National\ 16}$, which is the Case-Shiller National 16 city house price return series.

**Table 7**
**Housing Asset Pricing**
**Case-Shiller City Augmented Models**

$$R_{HPI} = \alpha_0 + \beta R_m + x'\delta + \varepsilon$$

| | (1) | (2) | (3) |
|---|---|---|---|
| $R_{CS\ national\ 10}$ | 0.662 | 0.666 | 0.655 |
| | (13) | (14) | (13) |
| SMB | 0.149 | 0.127 | 0.094 |
| | (4) | (1) | (1) |
| Momentum | -0.006 | -0.001 | 0.000 |
| | (1) | (1) | (1) |
| $s^2$ | 0.046 | 0.052 | 0.052 |
| | (5) | (5) | (5) |
| $\Delta$Emp | 0.021 | -0.007 | 0.001 |
| | (2) | (2) | (2) |
| Afford | | 1.61 | 1.31 |
| | | (4) | (4) |
| $\Delta$Forc | | | 0.001 |
| | | | (1) |
| **Distribution of CAPM $\beta$** | | | |
| mean | 0.662 | 0.666 | 0.655 |
| min | -0.081 | -0.122 | -0.124 |
| %50Q | 0.572 | 0.564 | 0.571 |
| max | 1.48 | 1.45 | 1.46 |
| **Distribution of $R^2$** | | | |
| mean | 0.617 | 0.661 | 0.667 |
| min | 0.243 | 0.32 | 0.327 |
| %50Q | 0.639 | 0.679 | 0.68 |
| max | 0.895 | 0.898 | 0.9 |

The mean coefficient values for variables of the models listed are presented for the 10 Case-Shiller MSAs. The number of MSAs from the sample with significant coefficients at the 5% level follows in parentheses. Summary details of the distribution of model betas and the distribution of $R^2$ follow. All models include an unreported constant. The timeframe is between 1990 and 2007 using quarterly data. The variables are defined as in Table 1 except for $R_{HPI}$ which here is the Case Shiller housing return series for each MSA, and $R_{CS\ National\ 16}$, which is the Case-Shiller National 10 MSA house price return series.

**Appendix Table 1**
**Investment Model (1) results for Individual MSAs**

| MSA | $R_{OFHEO}$ | | | $(R_{HPI}$ |
|---|---|---|---|---|
| | ß | SE (ß) | $R^2$ | mean |
| Akron, OH | 0.280 | 0.134 | 0.047 | 1.0572 |
| Albany-Schenectady-Troy, NY | 1.734 | 0.221 | 0.409 | 1.339 |
| Albuquerque, NM | 0.574 | 0.157 | 0.130 | 1.1294 |
| Allentown-Bethlehem-Easton, PA-NJ | 1.508 | 0.198 | 0.395 | 1.1506 |
| Amarillo, TX | 0.124 | 0.264 | 0.003 | 1.416 |
| Anchorage, AK | 0.417 | 0.536 | 0.007 | 1.3434 |
| Atlanta-Sandy Springs-Marietta, GA | 0.568 | 0.098 | 0.275 | 1.1117 |
| Atlantic City-Hammonton, NJ | 1.950 | 0.205 | 0.504 | 1.0665 |
| Augusta-Richmond County, GA-SC | 0.633 | 0.174 | 0.130 | 1.0458 |
| Austin-Round Rock, TX | 0.307 | 0.428 | 0.006 | 0.8995 |
| Bakersfield, CA | 2.358 | 0.228 | 0.545 | 1.299 |
| Barnstable Town, MA | 1.932 | 0.271 | 0.364 | 0.8431 |
| Baton Rouge, LA | 0.103 | 0.175 | 0.004 | 0.9941 |
| Beaumont-Port Arthur, TX | 0.056 | 0.200 | 0.001 | 1.0793 |
| Bellingham, WA | 0.721 | 0.277 | 0.071 | 1.1096 |
| Binghamton, NY | 1.064 | 0.332 | 0.104 | 0.762 |
| Birmingham-Hoover, AL | 0.473 | 0.107 | 0.179 | 1.0487 |
| Bloomington-Normal, IL | 0.036 | 0.142 | 0.001 | 0.9036 |
| Boise City-Nampa, ID | 0.714 | 0.233 | 0.095 | 0.8777 |
| Boston-Quincy, MA | 1.684 | 0.217 | 0.403 | 1.114 |
| Buffalo-Niagara Falls, NY | 0.702 | 0.166 | 0.168 | 0.8782 |
| Canton-Massillon, OH | 0.110 | 0.141 | 0.007 | 0.6668 |
| Casper, WY | -0.052 | 0.648 | 0.000 | 0.92 |
| Cedar Rapids, IA | 0.043 | 0.132 | 0.001 | 1.1479 |
| Charleston-North Charleston-Summerville, SC | 1.109 | 0.202 | 0.253 | 0.7596 |
| Charlotte-Gastonia-Concord, NC-SC | 0.207 | 0.093 | 0.052 | 0.8176 |
| Chattanooga, TN-GA | 0.473 | 0.140 | 0.114 | 0.8927 |
| Cheyenne, WY | 0.228 | 0.317 | 0.006 | 0.9769 |
| Chicago-Naperville-Joliet, IL | 0.818 | 0.074 | 0.578 | 0.9821 |
| Chico, CA | 1.695 | 0.240 | 0.360 | 0.7407 |
| Cincinnati-Middletown, OH-KY-IN | 0.301 | 0.061 | 0.214 | 0.7391 |
| Cleveland-Elyria-Mentor, OH | 0.270 | 0.099 | 0.077 | 0.9966 |
| Colorado Springs, CO | 0.304 | 0.160 | 0.039 | 1.0912 |
| Columbia, SC | 0.516 | 0.120 | 0.172 | 0.8038 |
| Columbus, OH | 0.278 | 0.072 | 0.145 | 0.8021 |
| Corpus Christi, TX | 0.541 | 0.248 | 0.051 | 0.7072 |
| Dallas-Plano-Irving, TX | 0.435 | 0.140 | 0.098 | 1.3007 |
| **Davenport-Moline-Rock Island, IA-IL** | **0.015** | **0.140** | **0.000** | **0.7276** |
| **Dayton, OH** | **0.221** | **0.097** | **0.055** | **0.7095** |
| **Deltona-Daytona Beach-Ormond Beach, FL** | **1.998** | **0.184** | **0.569** | **0.7694** |
| **Denver-Aurora, CO** | **0.103** | **0.170** | **0.004** | **1.2064** |
| **Des Moines-West Des Moines, IA** | **0.238** | **0.110** | **0.050** | **0.9168** |
| **Detroit-Livonia-Dearborn, MI** | **0.327** | **0.159** | **0.046** | **1.147** |
| **Eau Claire, WI** | **0.230** | **0.245** | **0.010** | **0.9843** |
| **El Paso, TX** | **0.646** | **0.198** | **0.107** | **1.0644** |
| **Elkhart-Goshen, IN** | **0.275** | **0.200** | **0.021** | **1.0589** |
| **Eugene-Springfield, OR** | **0.438** | **0.252** | **0.033** | **0.9337** |
| **Evansville, IN-KY** | **0.190** | **0.136** | **0.022** | **1.0188** |
| **Fayetteville-Springdale-Rogers, AR-MO** | **0.907** | **0.249** | **0.129** | **0.8549** |
| **Fort Collins-Loveland, CO** | **-0.043** | **0.197** | **0.001** | **1.1658** |
| **Fort Wayne, IN** | **0.381** | **0.120** | **0.101** | **0.6735** |
| **Fresno, CA** | **2.022** | **0.221** | **0.484** | **1.0541** |
| **Grand Junction, CO** | **0.229** | **0.392** | **0.004** | **0.9354** |
| **Grand Rapids-Wyoming, MI** | **0.232** | **0.114** | **0.044** | **0.8329** |
| **Greensboro-High Point, NC** | **0.345** | **0.128** | **0.075** | **1.2584** |
| **Greenville-Mouldin-Easley, SC** | **0.093** | **0.125** | **0.006** | **0.8208** |
| **Harrisburg-Carlisle, PA** | **0.756** | **0.152** | **0.218** | **0.9429** |
| **Hartford-West Hartford-East Hartford, CT** | **1.782** | **0.227** | **0.408** | **1.0086** |
| **Honolulu, HI** | **1.528** | **0.317** | **0.207** | **1.0602** |
| **Houston-Sugar Land-Baytown, TX** | **0.350** | **0.173** | **0.044** | **1.1855** |
| **Huntsville, AL** | **0.538** | **0.124** | **0.175** | **1.0944** |
| **Indianapolis-Carmel, IN** | **0.213** | **0.075** | **0.082** | **0.6723** |
| **Jackson, MS** | **0.458** | **0.176** | **0.071** | **1.3117** |
| **Jacksonville, FL** | **1.318** | **0.133** | **0.526** | **0.8367** |
| **Janesville, WI** | **-0.014** | **0.174** | **0.000** | **0.7472** |
| **Kalamazoo-Portage, MI** | **0.243** | **0.159** | **0.026** | **0.9528** |
| **Kansas City, MO-KS** | **0.432** | **0.073** | **0.282** | **0.8275** |
| **Knoxville, TN** | **0.461** | **0.150** | **0.097** | **0.9297** |
| **La Crosse, WI-MN** | **0.346** | **0.165** | **0.047** | **1.0272** |
| **Lafayette, LA** | **0.066** | **0.341** | **0.000** | **0.8962** |
| **Lancaster, PA** | **0.862** | **0.123** | **0.357** | **0.9361** |
| **Lansing-East Lansing, MI** | **0.441** | **0.113** | **0.146** | **0.8255** |
| **Las Cruces, NM** | **0.927** | **0.241** | **0.142** | **0.7691** |
| **Las Vegas-Paradise, NV** | **1.862** | **0.227** | **0.430** | **1.1377** |
| **Lexington-Fayette, KY** | **0.397** | **0.100** | **0.150** | **0.9061** |
| **Lima, OH** | **0.589** | **0.242** | **0.062** | **0.9898** |
| **Lincoln, NE** | **0.127** | **0.131** | **0.010** | **0.9559** |
| **Little Rock-North Little Rock-Conway, AR** | **0.438** | **0.151** | **0.086** | **1.2105** |
| **Longview, TX** | **0.280** | **0.305** | **0.009** | **0.9554** |
| **Los Angeles-Long Beach-Glendale, CA** | **2.573** | **0.213** | **0.621** | **0.7647** |
| **Louisville-Jefferson County, KY-IN** | **0.169** | **0.084** | **0.043** | **0.7858** |
| **Lubbock, TX** | **0.491** | **0.261** | **0.038** | **0.9269** |
| **Macon, GA** | **0.442** | **0.191** | **0.057** | **0.9929** |
| **Madison, WI** | **0.227** | **0.148** | **0.026** | **0.9358** |
| **Mansfield, OH** | **0.220** | **0.281** | **0.007** | **0.8078** |
| **Medford, OR** | **1.252** | **0.256** | **0.212** | **1.0171** |
| **Memphis, TN-MS-AR** | **0.498** | **0.129** | **0.143** | **1.0581** |
| **Merced, CA** | **2.434** | **0.325** | **0.387** | **0.9529** |
| **Miami-Miami Beach-Kendall, FL** | **1.612** | **0.189** | **0.451** | **1.024** |
| **Milwaukee-Waukesha-West Allis, WI** | **0.513** | **0.094** | **0.250** | **1.0952** |
| **Minneapolis-St. Paul-Bloomington, MN-WI** | **0.843** | **0.126** | **0.334** | **1.2384** |
| **Mobile, AL** | **0.367** | **0.279** | **0.019** | **1.3253** |
| **Modesto, CA** | **2.610** | **0.238** | **0.575** | **1.1743** |
| **Monroe, LA** | **0.094** | **0.247** | **0.002** | **1.0984** |
| **Nashville-Davidson--Murfreesboro--Franklin, TN** | **0.338** | **0.111** | **0.094** | **1.2139** |
| **New Orleans-Metairie-Kenner, LA** | **0.521** | **0.199** | **0.072** | **1.292** |
| **New York-White Plains-Wayne, NY-NJ** | **1.965** | **0.170** | **0.600** | **1.1556** |
| **Odessa, TX** | **0.606** | **0.530** | **0.015** | **1.0838** |
| **Oklahoma City, OK** | **0.242** | **0.216** | **0.014** | **1.6593** |
| **Omaha-Council Bluffs, NE-IA** | **0.107** | **0.101** | **0.013** | **1.127** |
| **Orlando-Kissimmee, FL** | **1.873** | **0.154** | **0.625** | **1.0958** |
| **Pensacola-Ferry Pass-Brent, FL** | **1.323** | **0.218** | **0.294** | **0.9333** |
| **Peoria, IL** | **0.129** | **0.172** | **0.006** | **1.0466** |
| **Philadelphia, PA** | **1.595** | **0.131** | **0.626** | **1.0173** |
| **Phoenix-Mesa-Scottdale, AZ** | **1.870** | **0.199** | **0.498** | **1.047** |
| **Pittsburgh, PA** | **0.354** | **0.112** | **0.100** | **0.805** |
| **Portland-South Portland-Biddeford, ME** | **1.551** | **0.205** | **0.391** | **1.0064** |
| **Portland-Vancouver-Beaverton, OR-WA** | **0.287** | **0.178** | **0.028** | **1.2385** |
| **Provo-Orem, UT** | **-0.185** | **0.260** | **0.006** | **1.1399** |
| **Pueblo, CO** | **0.175** | **0.319** | **0.003** | **0.8566** |
| **Raleigh-Cary, NC** | **0.135** | **0.107** | **0.018** | **1.2643** |
| **Reading, PA** | **1.222** | **0.173** | **0.360** | **1.2323** |
| **Redding, CA** | **1.637** | **0.248** | **0.329** | **1.2162** |
| **Reno-Sparks, NV** | **1.725** | **0.210** | **0.432** | **1.4882** |
| **Richmond, VA** | **1.103** | **0.100** | **0.578** | **1.3934** |
| **Roanoke, VA** | **0.623** | **0.186** | **0.112** | **1.5619** |
| **Rochester, MN** | **0.456** | **0.161** | **0.083** | **1.2028** |
| **Rochester, NY** | **0.610** | **0.117** | **0.234** | **1.0864** |
| **Rockford, IL** | **0.289** | **0.089** | **0.107** | **1.0411** |
| **Sacramento-Arden-Arcade-Roseville, CA** | **2.128** | **0.237** | **0.475** | **1.0926** |
| **Saginaw-Saginaw Township North, MI** | **0.094** | **0.192** | **0.003** | **1.2592** |
| **Salinas, CA** | **2.446** | **0.223** | **0.575** | **1.2281** |
| **Salt Lake City, UT** | **-0.130** | **0.236** | **0.003** | **1.1049** |
| **San Antonio, TX** | **0.578** | **0.221** | **0.071** | **1.1072** |
| **San Diego-Carlsbad-San Marcos, CA** | **2.116** | **0.207** | **0.540** | **0.9676** |
| **San Francisco-San Mateo-Redwood City, CA** | **1.812** | **0.232** | **0.407** | **1.0826** |
| **San Luis Obispo-Paso Robles, CA** | **2.181** | **0.255** | **0.451** | **1.323** |
| **Santa Barbara-Santa Maria-Goleta, CA** | **2.534** | **0.235** | **0.567** | **1.5065** |
| **Savannah, GA** | **0.840** | **0.181** | **0.194** | **1.0317** |
| **Scranton-Wilkes-Barre, PA** | **0.986** | **0.295** | **0.112** | **1.1283** |
| **Seattle-Bellevue-Everett, WA** | **0.636** | **0.215** | **0.090** | **1.2558** |
| **Shreveport-Bossier City, LA** | **0.384** | **0.216** | **0.034** | **0.8385** |
| **South Bend-Mishawaka, IN-MI** | **0.428** | **0.143** | **0.092** | **1.0376** |
| **Spokane, WA** | **0.421** | **0.222** | **0.039** | **1.1786** |
| **Springfield, IL** | **0.262** | **0.166** | **0.027** | **1.1349** |
| **Springfield, MA** | **1.793** | **0.203** | **0.467** | **1.2382** |
| **Springfield, MO** | **0.274** | **0.153** | **0.035** | **1.1083** |
| **St. Louis, MO-IL** | **0.646** | **0.064** | **0.533** | **0.7992** |
| **Stockton, CA** | **2.607** | **0.213** | **0.628** | **0.9443** |
| **Syracuse, NY** | **0.985** | **0.159** | **0.303** | **0.8166** |
| **Tallahassee, FL** | **1.070** | **0.188** | **0.266** | **0.8829** |
| **Tampa-St. Petersburg-Clearwater, FL** | **1.775** | **0.133** | **0.668** | **1.4276** |
| **Toledo, OH** | **0.311** | **0.110** | **0.082** | **1.2064** |
| **Topeka, KS** | **0.351** | **0.211** | **0.030** | **1.2281** |
| **Tucson, AZ** | **1.368** | **0.179** | **0.396** | **1.0605** |
| **Tulsa, OK** | **0.227** | **0.174** | **0.019** | **1.2278** |
| **Tyler, TX** | **0.429** | **0.413** | **0.012** | **1.7807** |
| **Visalia-Porterville, CA** | **1.990** | **0.225** | **0.469** | **1.0256** |
| **Washington-Arlington-Alexandria, DC-VA-MD-WV** | **2.105** | **0.128** | **0.752** | **0.941** |
| **Waterloo-Cedar Falls, IA** | **0.454** | **0.386** | **0.015** | **0.7847** |
| **York-Hanover, PA** | **1.018** | **0.157** | **0.320** | **0.8517** |

Market betas estimates from model (1) in Table 3 with the associated standard errors of betas and the $R^2$ are presented for each MSA. The mean return for each MSA is also presented. The timeframe is between 1985 and 2007 using quarterly data.